\documentclass[prd,nofootinbib,preprint,superscriptaddress]{revtex4}
\pdfoutput=1
\usepackage[T1]{fontenc}
\usepackage{amsmath,amssymb}
\usepackage{epsfig}
\usepackage{graphicx}
\usepackage[usenames,dvipsnames]{color}
\usepackage{subfigure}
\usepackage{slashed}
\usepackage[colorlinks,citecolor=blue]{hyperref}
\usepackage{pdfpages}
\usepackage{color}

\begin{document}
\title{Low scale Dirac leptogenesis and dark matter with observable $\Delta N_{\rm eff}$}

\author{Devabrat Mahanta}
\email{devab176121007@iitg.ac.in}
\affiliation{Department of Physics, Indian Institute of Technology Guwahati, Assam 781039, India}

\author{Debasish Borah}
\email{dborah@iitg.ac.in}
\affiliation{Department of Physics, Indian Institute of Technology Guwahati, Assam 781039, India}

\begin{abstract}
We propose a gauged $B-L$ extension of the standard model (SM) where light neutrinos are of Dirac type by virtue of tiny Yukawa couplings. To achieve leptogenesis, we include additional heavy Majorana fermions without introducing any $B-L$ violation by two units. An additional scalar doublet with appropriate $B-L$ charge can allow such heavy fermion coupling with SM leptons so that out-of-equilibrium decay of the former can lead to generation of lepton asymmetry. Due to the $B-L$ gauge interactions of the decaying fermion, the criteria of successful Dirac leptogenesis can also constrain the gauge sector couplings. The same $B-L$ gauge sector parameter space can also be constrained from dark matter requirements if the latter is assumed to consist of SM singlet particles with non-zero $B-L$ charges, which also keep the model anomaly free. The same $B-L$ gauge interactions also lead to additional thermalised relativistic degrees of freedom $\Delta N_{\rm eff}$ from light Dirac neutrinos which are tightly constrained by Planck 2018 data. While there exists parameter space from the criteria of successful leptogenesis, dark matter and $\Delta N_{\rm eff}$ even after incorporating the latest collider bounds, the currently allowed parameter space can be probed by future measurements of $\Delta N_{\rm eff}$. 

\end{abstract}

\maketitle

\section{Introduction}
The observed matter-antimatter asymmetry of the universe has been a longstanding puzzle \cite{Zyla:2020zbs, Aghanim:2018eyx}. This asymmetry is often quoted as a ratio of excess of baryons over antibaryons to photons. According to the latest data from Planck satellite it is given as 
\cite{Aghanim:2018eyx} 
\begin{equation}
\eta_B = \frac{n_{B}-n_{\bar{B}}}{n_{\gamma}} = 6.1 \times 10^{-10}.
\label{etaBobs}
\end{equation}
This cosmic microwave background (CMB) based measurements of baryon asymmetry matches very well with the bing bang nucleosynthesis (BBN) too. Since the universe is expected to start in a baryon symmetric manner and any initial asymmetry is expected to be diluted due to accelerated expansion during inflation, the observed asymmetry calls for a dynamical explanation. Such a dynamical process requires to satisfy certain conditions, given by Sakharov \cite{Sakharov:1967dj} as (1) baryon number (B) violation, (2) C and CP violation and (3) departure from thermal equilibrium, all of which can not be satisfied in required amount in the standard model (SM) of particle physics and considering an expanding Friedmann-Lemaitre-Robertson-Walker (FLRW) universe. This has led to many beyond standard model (BSM) proposals where out-of-equilibrium and B, C, CP violating decays of newly introduced heavy particles can lead to generation of the baryon asymmetry of the universe (BAU) \cite{Weinberg:1979bt, Kolb:1979qa}. One interesting way to achieve baryogenesis is through leptogenesis where an asymmetry in lepton sector is generated (through lepton number (L) violating interactions) first which later gets converted into baryon asymmetry through $(B+L)$-violating EW sphaleron transitions~\cite{Kuzmin:1985mm}. First proposed by Fukugita and Yanagida more than thirty years back \cite{Fukugita:1986hr}, leptogenesis has become very popular as the asymmetry can be generated by the same fields which also take part in seesaw origin of light neutrino masses \cite{Minkowski:1977sc, Mohapatra:1979ia, Yanagida:1979as, GellMann:1980vs, Glashow:1979nm, Schechter:1980gr}. Origin of light neutrino masses and mixing \cite{Zyla:2020zbs} is another observed phenomena which SM fails to explain and typical seesaw models can explain both light neutrino masses and baryon asymmetry through leptogenesis.

It is worth mentioning that the typical seesaw models give rise to Majorana light neutrinos as they violate lepton number by 2 units. However, the nature of light neutrinos: Dirac or Majorana, is not yet known. While neutrino oscillation experiments can not settle this issue, there are other experiments like the ones looking for neutrinoless double beta decay ($0\nu \beta \beta$) \cite{Rodejohann:2011mu}, a promising signature of Majorana neutrinos. However, there have been no such observations yet which can confirm Majorana nature of light neutrinos. While null result at $0\nu \beta \beta$ does not necessarily rule out Majorana nature of light neutrinos, it has led to growing interests in Dirac neutrino models in the last few years. For different seesaw realisations to generate light Dirac neutrino masses, please see \cite{Babu:1988yq, Peltoniemi:1992ss, Chulia:2016ngi, Aranda:2013gga, Chen:2015jta, Ma:2015mjd, Reig:2016ewy, Wang:2016lve, Wang:2017mcy, Wang:2006jy, Gabriel:2006ns, Davidson:2009ha, Davidson:2010sf, Bonilla:2016zef, Farzan:2012sa, Bonilla:2016diq, Ma:2016mwh, Ma:2017kgb, Borah:2016lrl, Borah:2016zbd, Borah:2016hqn, Borah:2017leo, CentellesChulia:2017koy, Bonilla:2017ekt, Memenga:2013vc, Borah:2017dmk, CentellesChulia:2018gwr, CentellesChulia:2018bkz, Han:2018zcn, Borah:2018gjk, Borah:2018nvu, CentellesChulia:2019xky,Jana:2019mgj, Borah:2019bdi, Dasgupta:2019rmf, Correia:2019vbn, Ma:2019byo, Ma:2019iwj, Baek:2019wdn, Saad:2019bqf, Jana:2019mez, Nanda:2019nqy} and references therein. While Dirac neutrinos may imply conservation of lepton number and absence of a viable leptogenesis mechanism, there are interesting ways to circumvent this. As proposed in \cite{Dick:1999je, Murayama:2002je}, one can have successful leptogenesis even with light Dirac neutrino scenarios where total lepton number or $B-L$ is conserved just like in the SM. Popularly known as Dirac leptogenesis scenario, it involves the creation of an equal and opposite amount of lepton asymmetry in left handed and right handed neutrino sectors followed by the conversion of left sector asymmetry into baryon asymmetry via electroweak sphalerons. The lepton asymmetries left and right handed sectors are prevented from equilibration due to the tiny effective Dirac Yukawa couplings. For similar works in the context of different models, please see \cite{Boz:2004ga, Thomas:2005rs, Thomas:2006gr, Cerdeno:2006ha, Gu:2006dc, Gu:2007mc, Chun:2008pg, Bechinger:2009qk, Chen:2011sb, Choi:2012ba, Borah:2016zbd, Gu:2016hxh, Narendra:2017uxl}. In a few other works \cite{Heeck:2013vha, Gu:2019yvw} light Dirac neutrinos with $B-L$ violation was considered in such a way that the latter does not give any Majorana mass for light neutrinos. We find this interesting due to the role $B-L$ gauge symmetry or the corresponding gauge boson after spontaneous symmetry breaking can play in phenomenology and other experimental signatures. It should be noted that leptogenesis with light Dirac neutrinos have been referred to as Dirac leptogenesis, Dirac neutrinogenesis among others in the above-mentioned works depending upon the particular realisations. Here, we simply call it Dirac leptogenesis for simplicity.

Motivated by these, we consider a gauged $B-L$ model with light Dirac neutrinos and lepton number violating heavy chiral fermions whose out-of-equilibrium decay into leptons lead to generation of lepton asymmetry. In contrast with earlier works \cite{Heeck:2013vha, Gu:2019yvw}, here lepton asymmetry is generated by decay processes similar to usual type I seesaw \cite{Minkowski:1977sc, Mohapatra:1979ia, Yanagida:1979as, GellMann:1980vs, Glashow:1979nm, Schechter:1980gr} leptogenesis except for the fact that there are no $\Delta (B-L) = 2$ processes ensuring Dirac nature of light neutrinos. The $B-L$ gauge boson plays important role in leptogenesis, specially in wash-out processes. The same $B-L$ gauge interactions also lead to thermalisation of right handed neutrinos contributing to relativistic degrees of freedom $N_{\rm eff}$ which is tightly constrained from Planck 2018 data. We also consider $B-L$ portal fermion singlet dark matter (DM) whose relic density depends crucially upon its annihilation channels mediated by $B-L$ gauge bosons. DM comprises approximately 26\% of present universe's energy density \cite{Zyla:2020zbs, Aghanim:2018eyx}. We consider a typical weakly interacting massive particle (WIMP) DM which gets thermally produced in the early universe from the standard bath by virtue of $B-L$ gauge interaction followed by freeze-out leaving a relic. Thus, the requirements of successful leptogenesis and correct dark matter relic abundance without overproducing $N_{\rm eff}$ tightly constrain the $B-L$ gauge sector parameter space, in addition to the constraints from collider experiments. While collider bounds rule out some parts of the parameter space, the Planck 2018 bound on $N_{\rm eff}$ rules out some additional parameter space consistent with leptogenesis and dark matter requirements. We show that future CMB experiments with much more sensitivity to $N_{\rm eff}$ will be able to probe the entire parameter space currently favoured from low scale leptogenesis and dark matter requirements. We also show that the fermion singlet DM candidates with exotic $B-L$ charges keep the model anomaly free, adding another motivation to consider DM in this setup. We find one minimal framework of this type where two Dirac fermion DM candidates are sufficient to cancel the anomalies. This allowed parameter space consistent with DM, leptogenesis and $N_{\rm eff}$ criteria while satisfying collider bounds can still be probed at near future CMB experiments, keeping the model very predictive and testable.

This paper is organised as follows. In section \ref{sec1}, we discuss our basic framework in the context of Dirac leptogenesis and $N_{\rm eff}$. In section \ref{sec2} we discuss our UV complete anomaly free model and incorporate the relevant DM phenomenology followed by summary of the results. Finally we conclude in section \ref{sec3}.

\section{The Basic Framework}
\label{sec1}
In this section, we consider a minimal setup based on gauge $B-L$ extension of the SM to highlight our key findings. Gauged $B-L$ extension of the SM \cite{Davidson:1978pm, Mohapatra:1980qe, Marshak:1979fm, Masiero:1982fi, Mohapatra:1982xz, Buchmuller:1991ce} has been a popular framework studied in the context of neutrino mass, leptogenesis for a long time. In addition to the SM fermion content with usual gauge charges, we consider three right handed neutrinos $\nu_R$ having $B-L$ charge -1 which form massive Dirac neutrinos with $\nu_L$ after electroweak symmetry breaking (EWSB). The scalar content is chosen in a way that prevents $\nu_R$ from acquiring Majorana masses. To realise leptogenesis, we introduce two heavy singlet fermions $N_R$ having $B-L$ charge $n_1 \neq \pm 1$ so that generating Majorana mass term of $N_R$ (generated by singlet scalar $\phi_1$) does not lead to Majorana mass of $\nu_R$. Since lepton doublet $\ell_L$ and $N_{R}$ have different $B-L$ charges, we introduce another scalar doublet $\eta$ with appropriate non-zero $B-L$ charge so that the required Yukawa coupling can be realised. The neutral component of this new scalar doublet does not acquire any vacuum expectation value (VEV), a requirement to ensure that light neutrinos do not receive any Majorana mass contribution via type I seesaw. Note that, this particle content will not lead to an anomaly free gauged $B-L$ model. We will later show how additional chiral fermions with non-trivial $B-L$ charges can lead to vanishing anomalies while playing the role of dominant dark matter component of the universe at the same time.
With the particle content mentioned above, the relevant Yukawa Lagrangian is 
\begin{align}
\mathcal{L}_Y \supset Y_D \bar{L} \tilde{H} \nu_R + Y_{\eta} \bar{L} \tilde{\eta} N_R + Y_N \phi_1 N_R N_R+{\rm h.c.}
\end{align}
Thus, light Dirac neutrino masses arise purely from the Yukawa coupling with the SM Higgs. While it requires fine-tuned Yukawa coupling to generate sub-eV Dirac neutrino masses, we do not pursue dynamical realisations of such small Yukawa couplings in the minimal models discussed here. Details of such Dirac neutrino models can be found in \cite{Babu:1988yq, Peltoniemi:1992ss, Chulia:2016ngi, Aranda:2013gga, Chen:2015jta, Ma:2015mjd, Reig:2016ewy, Wang:2016lve, Wang:2017mcy, Wang:2006jy, Gabriel:2006ns, Davidson:2009ha, Davidson:2010sf, Bonilla:2016zef, Farzan:2012sa, Bonilla:2016diq, Ma:2016mwh, Ma:2017kgb, Borah:2016lrl, Borah:2016zbd, Borah:2016hqn, Borah:2017leo, CentellesChulia:2017koy, Bonilla:2017ekt, Memenga:2013vc, Borah:2017dmk, CentellesChulia:2018gwr, CentellesChulia:2018bkz, Han:2018zcn, Borah:2018gjk, Borah:2018nvu, CentellesChulia:2019xky,Jana:2019mgj, Borah:2019bdi, Dasgupta:2019rmf, Correia:2019vbn, Ma:2019byo, Ma:2019iwj, Baek:2019wdn, Saad:2019bqf, Jana:2019mez, Nanda:2019nqy}.

The gauge invariant scalar interactions can be written as  
\begin{align}
\mathcal{L}_{scalar} &= \left({D_{H}}_{\mu} H \right)^\dagger
\left({D_{H}}^{\mu} H \right) + \left({D_{\eta}}_{\mu} \eta \right)^\dagger \left({D_{\eta}}^{\mu} \eta \right)+\left({D_{\phi}}_{\mu} \phi_1 \right)^\dagger \left({D_{\phi}}^{\mu} \phi_1 \right) -\mu^2_H \lvert H \rvert^2  \nonumber \\ 
& + \lambda_H \lvert H \rvert^4 + \left( \mu^2_{\eta} \lvert \eta \rvert^2 + \lambda_{\eta} \lvert \eta \rvert^4 \right) +\lambda_{H\eta} (\eta^{\dagger} \eta) (H^{\dagger} H) + \lambda^{\prime}_{H\eta} (\eta^{\dagger} H) (H^{\dagger} \eta) \nonumber \\
& -\mu_{\phi_{1}}^{2} \lvert \phi_{1} \rvert^{2}+\lambda_{\phi_{1}}\lvert \phi_{1} \rvert^{4}+\lambda_{H\phi_{1}}(H^{\dagger}H)(\phi_{1}^{\dagger}\phi_{1}) + \lambda_{\eta \phi_{1}} (\eta^{\dagger}\eta)(\phi_{1}^{\dagger}\phi_{1})
\label{scalar:lag}
\end{align}
where $\rm{{D_{H}}^{\mu}}$, $\rm{{D_{\eta}}^{\mu}}$ and $\rm{{D_{\phi}}^{\mu}}$ denote the covariant derivatives for the scalar doublets H, $\rm{\eta}$ and scalar singlets ${\rm\phi_i}$ respectively and can be written as 

\begin{eqnarray}
{D_{H}}_{\mu}\,H &=& \left(\partial_{\mu} + i\,\dfrac{g}{2}\,\sigma_a\,W^a_{\mu}
+ i\,\dfrac{g^\prime}{2}\,B_{\mu}\right)H \,, \nonumber \\
{D_{\eta}}_{\mu}\,\eta &=& \left(\partial_{\mu} + i\,\dfrac{g}{2}\,\sigma_a\,W^a_{\mu}
+ i\,\dfrac{g^\prime}{2}\,B_{\mu} + i\,g_{BL}\,n_{\eta}
{Z_{BL}}_{\mu}\right)\eta \,, \nonumber \\
{D_{\phi}}_{\mu}\,\phi_1 &=& \left(\partial_{\mu} + i\,g_{BL}\,n_{\phi_1}
{Z_{BL}}_{\mu}\right)\phi_1\,.
\end{eqnarray}
Here ${\rm g_{BL}}$ is the new gauge coupling and ${\rm n_{\eta}}$ and ${\rm n_{\phi_1}}$ are the charges under ${\rm U(1)_{B-L}}$ for ${\rm \eta}$ and ${\rm \phi_1}$ respectively. The kinetic term of the $B-L$ gauge boson can be written as 
\begin{equation}
 \mathcal{L}_{\rm kin} \supset -\frac{1}{4} B'_{\alpha \beta} B'^{\alpha \beta}
\end{equation}
where $B'^{\alpha \beta} = \partial^{\alpha} Z^{\beta}_{BL}-\partial^{\beta} Z^{\alpha}_{BL}$ is the corresponding field strength tensor. The symmetry of the model also allows kinetic mixing between $U(1)_Y$ of SM and $U(1)_{B-L}$ of the form $\epsilon B^{\alpha \beta} B'_{\alpha \beta}/2$ where $B^{\alpha \beta} = \partial^{\alpha} B^{\beta}-\partial^{\beta} B^{\alpha}$ and $\epsilon$ is the kinetic mixing parameter. We ignore such tree level mixing in this work. The mixing at one loop can be approximated as $\epsilon \approx g_{BL} g'/(16 \pi^2) $ \cite{Mambrini:2011dw}. As we will see while discussing our numerical results, our final allowed parameter space for $g_{BL}$ will correspond to tiny one-loop kinetic mixing (smaller than $\mathcal{O}(10^{-3})$) and has very little effect on the phenomenology discussed here. Therefore, we ignore such kinetic mixing in our work.

After both $B-L$ and electroweak symmetries get broken by the VEVs of H and $\phi_1$ the doublet and all three
singlets are given by

\begin{eqnarray}
H=\begin{pmatrix}H^+\\
\dfrac{h^{\prime} + v + i z}{\sqrt{2}}\end{pmatrix}\,,\,\,\,\,\,\,
\eta=\begin{pmatrix}\eta^+\\
\dfrac{\eta_R^{\prime} + i \eta_I^\prime}{\sqrt{2}}\end{pmatrix}\,,\,\,\,\,\,\,
\phi_1 = \dfrac{s^{\prime}_1 +u_i+ A^{\prime}_1}{\sqrt{2}}\,\,\,\,\,\,,
\label{H&phi_broken_phsae}
\end{eqnarray} 
The details of the scalar mass spectrum can be found in appendix \ref{scalar2}.

While tree level Majorana mass of light neutrinos is absent due to vanishing VEV of scalar doublet $\eta$, it is also necessary to ensure the absence or smallness of radiative Majorana neutrino masses. For example, at one loop level similar particle content can give rise to light neutrino masses via scotogenic fashion \cite{Ma:2006km}. However, such a contribution is absent in our setup. This is due to non-zero $B-L$ charge of scalar doublet $\eta$ as well as the choice of singlet scalars, preventing appropriate VEV insertions in the scalar lines of the one-loop diagrams. One can also notice it from the scalar Lagrangian above where the $ \eta \eta H^{\dagger} H^{\dagger}$, responsible for non-zero neutrino mass in scotogenic model \cite{Ma:2006km} remains absent.

\begin{figure}[h!]
\centering
\includegraphics[width=0.9\textwidth]{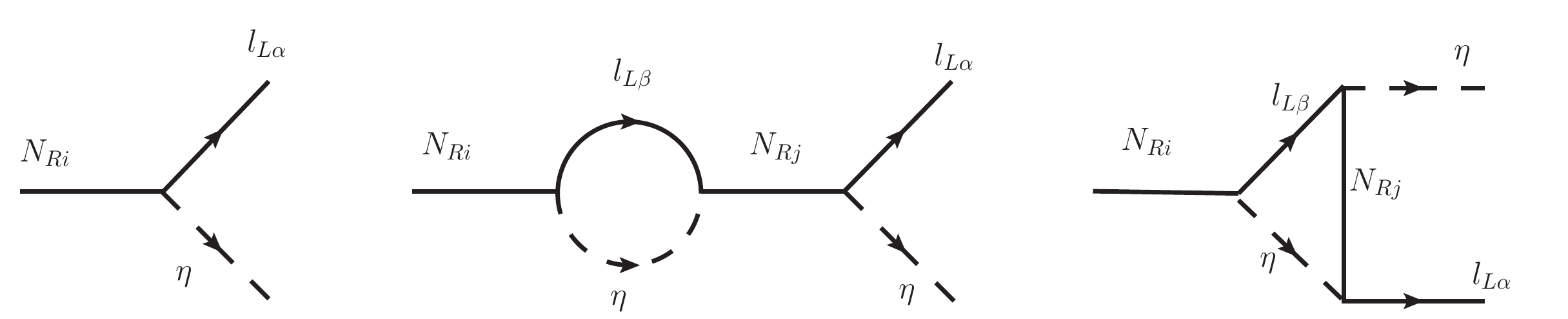}
\caption{Processes creating lepton asymmetry.}
\label{Fig:feyn}
\end{figure}

\subsection{Dirac leptogenesis}
As can be seen from the Yukawa Lagrangian mentioned above, the out-of-equilibrium decay of $N_R$ to lepton doublet and additional scalar doublet $\eta$ can generate lepton asymmetry in our model. Compared to usual Dirac leptogenesis scenarios, here the generation of lepton asymmetry is much simpler and only requires the steps followed in type I seesaw leptogenesis known for decades. The Feynman diagrams for this decay process, including one loop corrections are shown in figure \ref{Fig:feyn}.

The relevant Boltzmann equations can be written as

\begin{align}
\dfrac{dn_{N_{1}}}{dz} & = -D_{1}(n_{N_1}-n_{N_{1}}^{\rm eq})-\dfrac{s}{H(z)z} [ (n_{N_1}^2-(n_{N_{1}}^{\rm eq})^{2})\langle \sigma v \rangle_{N_{R_{1}}N_{R_{1}} \longrightarrow X X} ],       
\label{eq:5}
\end{align}
\begin{align}\label{eq:6}
        \dfrac{dn_{B-L}}{dz} & =-\epsilon_1 D_{1}(n_{N_{1}}-n_{N_{1}}^{\rm eq})-W^{\rm Total}n_{B-L} 
\end{align}
where $n_{N_1}$ denotes the comoving number density of $N_{R_1}$ (to be denoted as $N_1$ hereafter) and $n_{N_1}^{\rm eq}=\frac{z^2}{2}K_2(z)$ is its equilibrium number density (with $K_i(z)$ being the modified Bessel function of $i$-th kind). The quantity on the right hand side of the above equations
\begin{align}
D_1 \ \equiv \ \frac{\Gamma_{1}}{Hz} \ = \ K_{N_1} z \frac{K_1(z)}{K_2(z)}
\end{align}
 measures the total decay rate of $N_1$ with respect to the Hubble expansion rate, and similarly, $W^{\rm Total} \equiv \frac{\Gamma_{W}}{Hz}$ measures the total washout rate. Here, $K_{N_1}=\Gamma_1/H(z=1)$ is the decay parameter. In the second term on right hand side of equation \eqref{eq:5}, we consider $N_1$ annihilations into all SM fermions mediated by $Z_{BL}$ along with $N_1$ annihilations to pair of $Z_{BL}$ as well as $\eta$. Similarly, $n_{B-L}$ in equation \eqref{eq:6} denotes the comoving number density of $B-L$ generated from CP violating out-of-equilibrium decay of $N_1$. The CP asymmetry parameter on the right hand side of equation \eqref{eq:6} is defined as
\begin{equation}
\epsilon_{i} =\frac{\sum_{\alpha}\Gamma(N_{i}\rightarrow l_{\alpha}\eta)-\Gamma(N_{i}\rightarrow\bar{l_{\alpha}}\bar{\eta})}{\sum_{\alpha}\Gamma(N_{i}\rightarrow l_{\alpha}\eta)+\Gamma(N_{i}\rightarrow\bar{l_{\alpha}}\bar{\eta})}.
\label{epsilon1}
\end{equation} 

We consider a hierarchical limit of right handed neutrinos $N_R$ and hence only the lightest of them namely, $N_{R_1} \equiv N_1$ contribute dominantly to the generation of lepton asymmetry. In the Boltzmann equation of $N_{1}$ above, we have considered two types of dilution terms on the right hand side: one due to its decay into leptons and the other due to its annihilation into lighter particles. We consider the singlet scalar to be much heavier than the right handed neutrinos and hence the corresponding processes involving singlet scalars are sub-dominant compared to the ones mediated by $B-L$ gauge bosons. Such additional dilution terms for right handed neutrinos also appear in type I seesaw leptogenesis with Majorana light neutrinos and $B-L$ gauge symmetry, see \cite{Iso:2010mv, Okada:2012fs, Heeck:2016oda, Dev:2017xry} for details. In the Boltzmann equation for lepton asymmetry, in addition to the inverse decay, we consider the following scattering processes responsible for washing out the generated lepton asymmetry:
$$ l Z_{BL} \longrightarrow \eta N_{1}, l W^{\pm}(Z)\longleftarrow \eta N_{1}, \eta l \longleftarrow N_{1} Z_{BL}, l \eta \longleftarrow \bar{l} \eta^{*}, l N_{1} \longrightarrow \bar{l}N_{1}^{*}. $$
All the relevant cross sections and decay widths are given in Appendix \ref{washouts}. Note that, apart from the usual Yukawa or SM gauge coupling related processes, we also have washout processes involving $B-L$ gauge bosons. Since interactions involving $B-L$ gauge bosons can cause dilution of $N_{1}$ abundance as well as wash out the generated lepton asymmetry, one can tightly constrain the $B-L$ gauge sector couplings from the requirement of successful leptogenesis at low scale. Also, the Yukawa couplings involved in CP asymmetry generation are not related to light neutrino mass generation, unlike in vanilla leptogenesis scenario. Therefore, we do not have any strict lower bound on the scale of leptogenesis like the Davidson Ibarra bound of vanilla leptogenesis scenario \cite{Davidson:2002qv}.

\begin{figure}
\begin{center}
 \includegraphics[scale=.52]{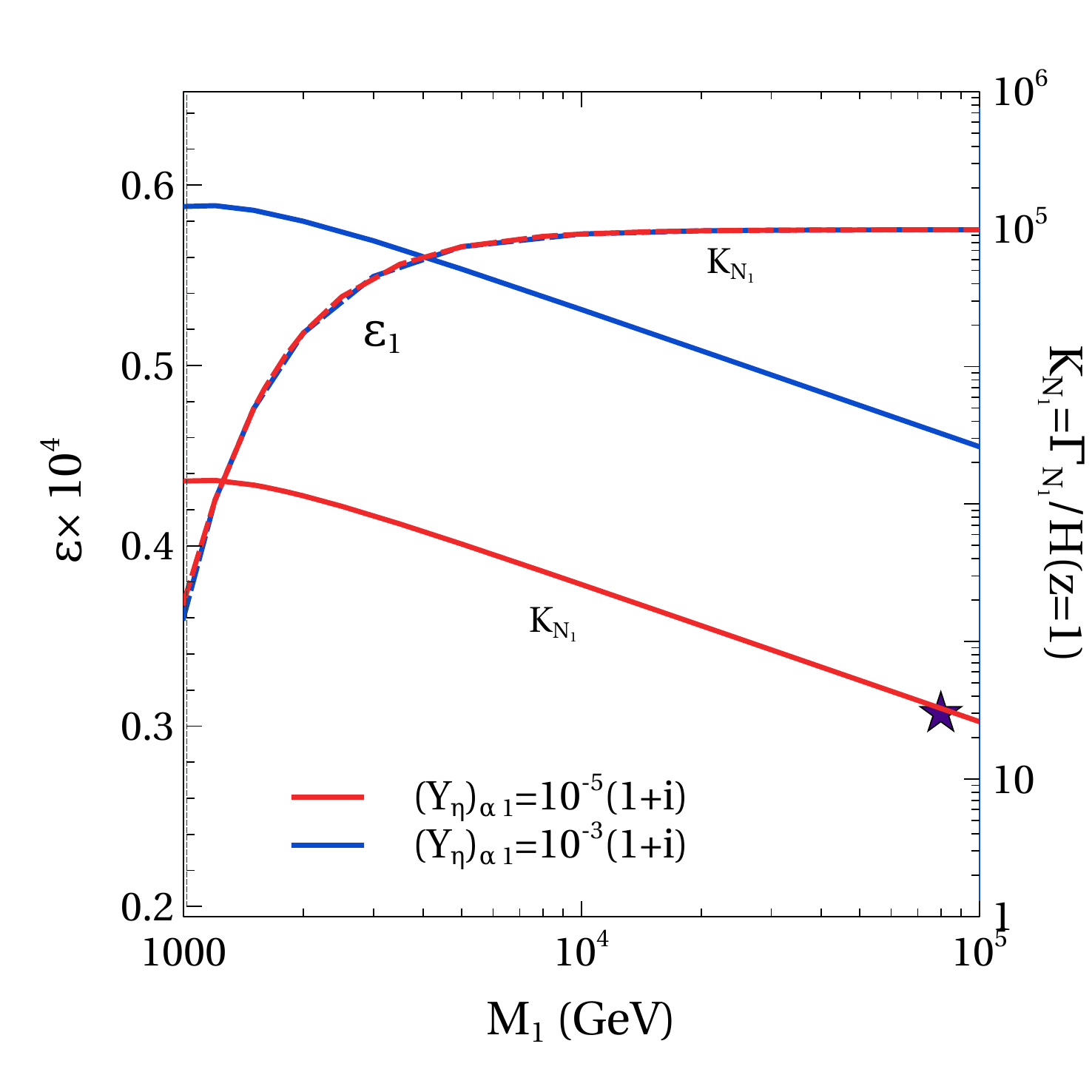}
 \includegraphics[scale=0.52]{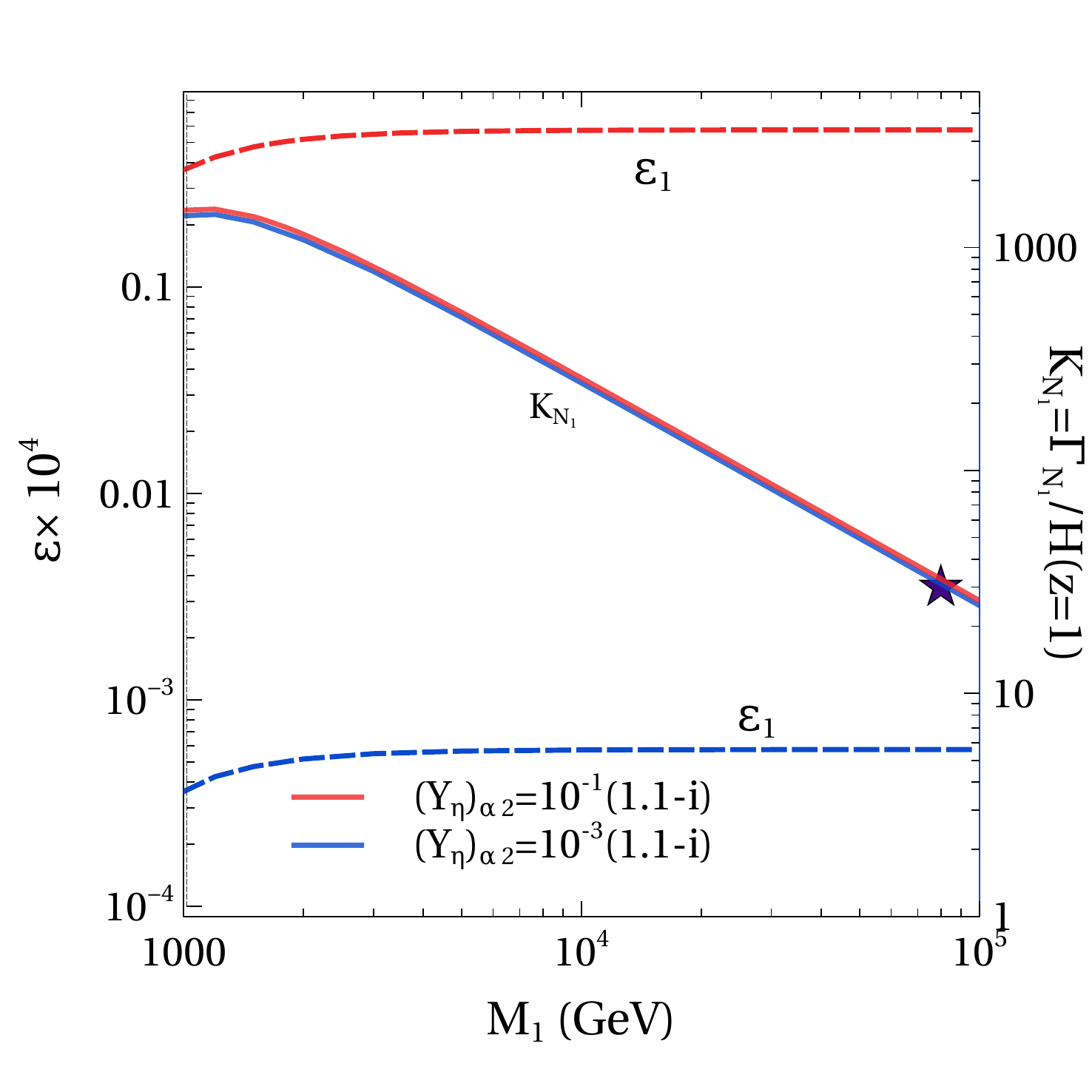}
 \caption{Variation of CP asymmetry parameter $\epsilon_1$ and the decay parameter $K_{N_{1}}$ with $M_{1}$ for different values of the $N_{1}$ Yukawa coupling (left panel) and for different values of the $N_{2}$ Yukawa coupling (Right panel). $K_{N_{1}}$ is shown by the solid lines and the $\epsilon_1$ is shown with the dashed lines. We fix the $(Y_{\eta})_{\alpha 2}=10^{-1}(1.1-i)$ for the left panel and $(Y_{\eta})_{\alpha 1}=10^{-5}(1+i)$ for the right panel. The points marked as stars correspond to successful leptogenesis scale and couplings.}
 \label{epsilon_kappa}
\end{center}
\end{figure}

\begin{figure}[h]
\begin{center}
\includegraphics[scale=.5]{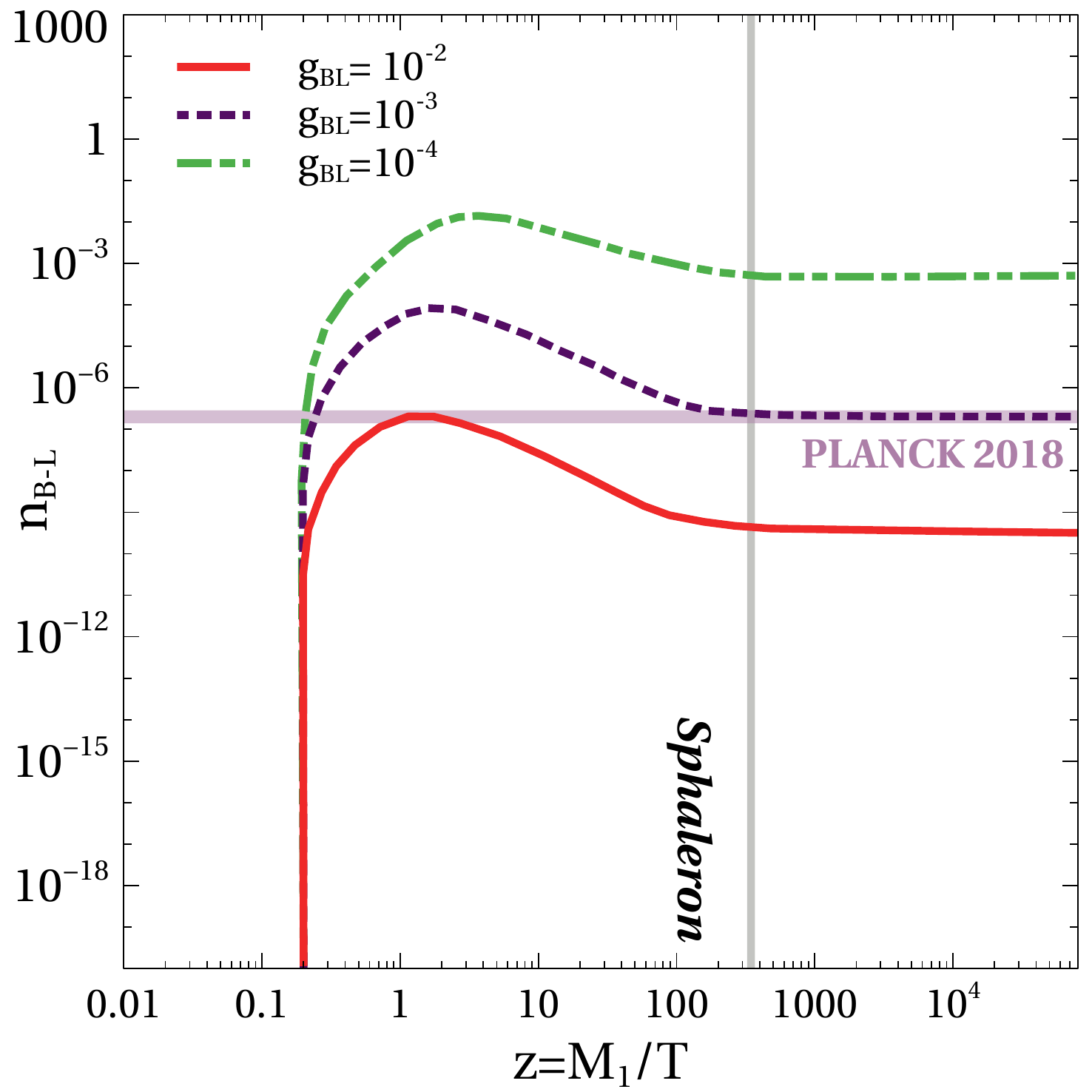}
\includegraphics[scale=.5]{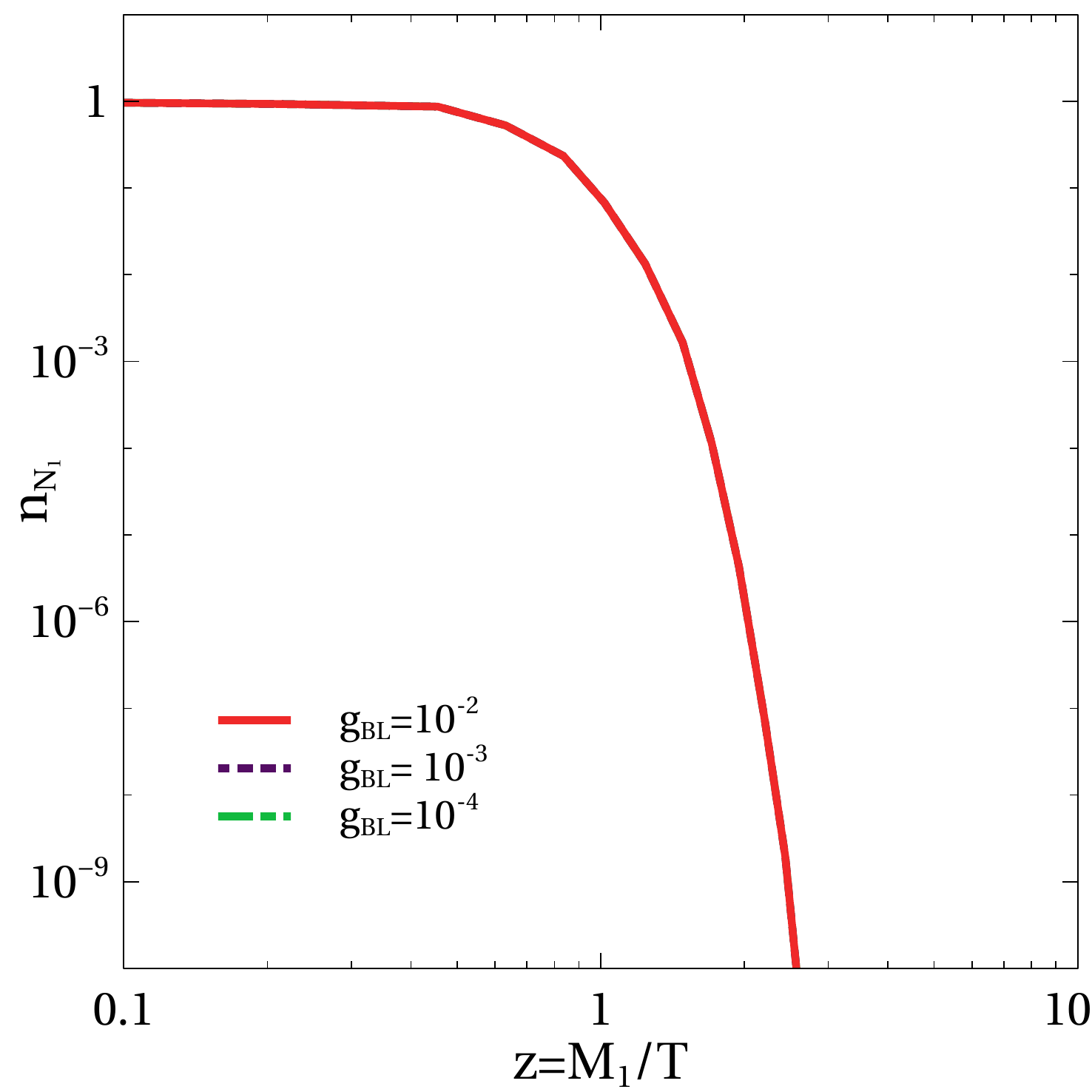}
\caption{Comoving density of B-L asymmetry (left panel) and comoving number density of $N_{R_{1}}$ (right panel) with $z=\dfrac{M_{1}}{T}$ for different $g_{BL}$. The Yukawa couplings relevant for Leptogenesis are taken to be $(Y_{\eta})_{\alpha 1}=10^{-5}(1+i)$ and $(Y_{\eta})_{\alpha 2}=10^{-1}(1.1-i)$. The other important parameters used are $M_{1}=45$ TeV, $M_{2}=450$ TeV, $m_{\eta}=5$ TeV and $M_{Z_{BL}}=4$ GeV. The horizontal line in left panel plot denotes the required $B-L$ asymmetry to generate for observed baryon asymmetry (Planck 2018) after sphaleron transition.}
\label{leptofig1}
\end{center}
\end{figure}

\begin{figure}[h]
\begin{center}
\includegraphics[scale=.5]{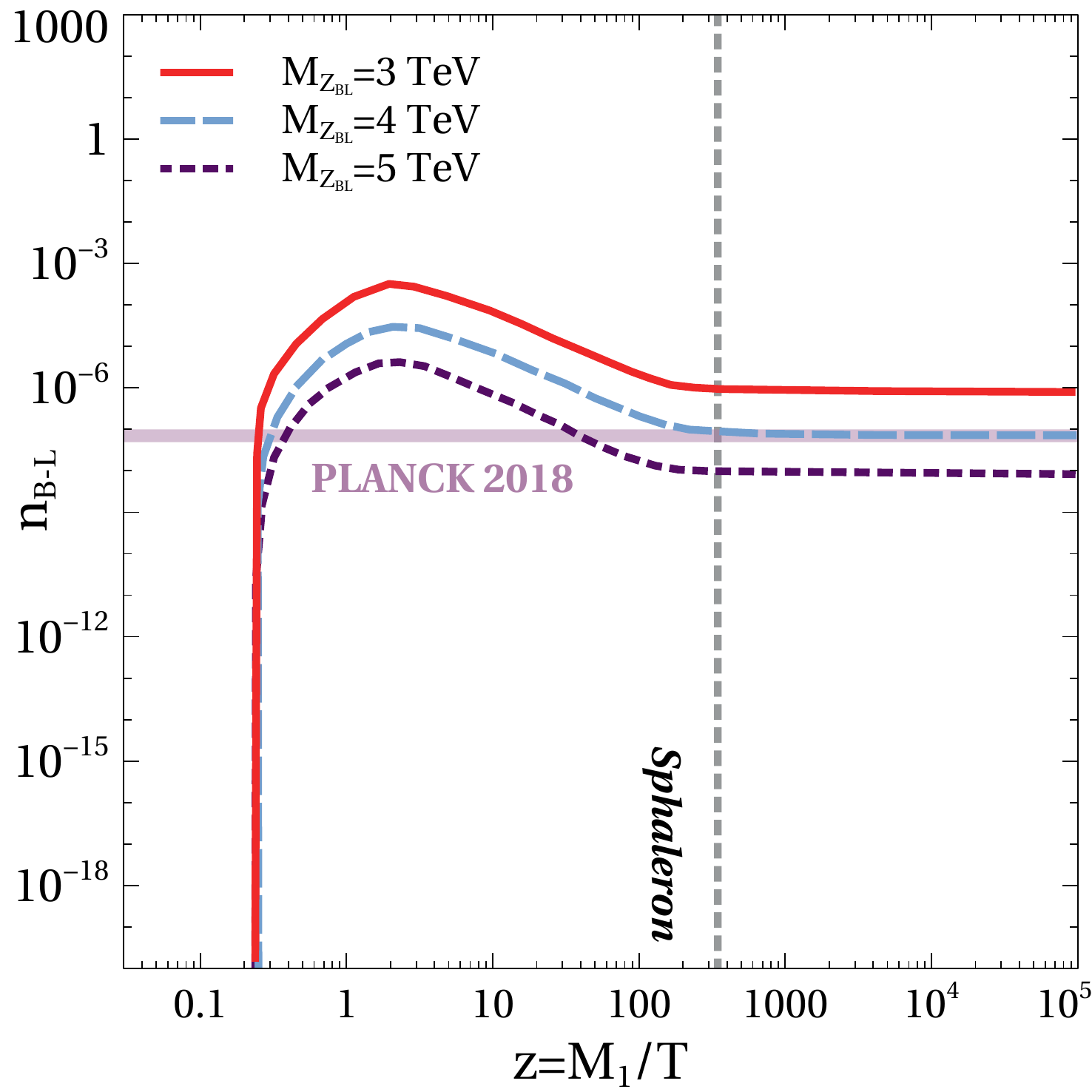}
\includegraphics[scale=.5]{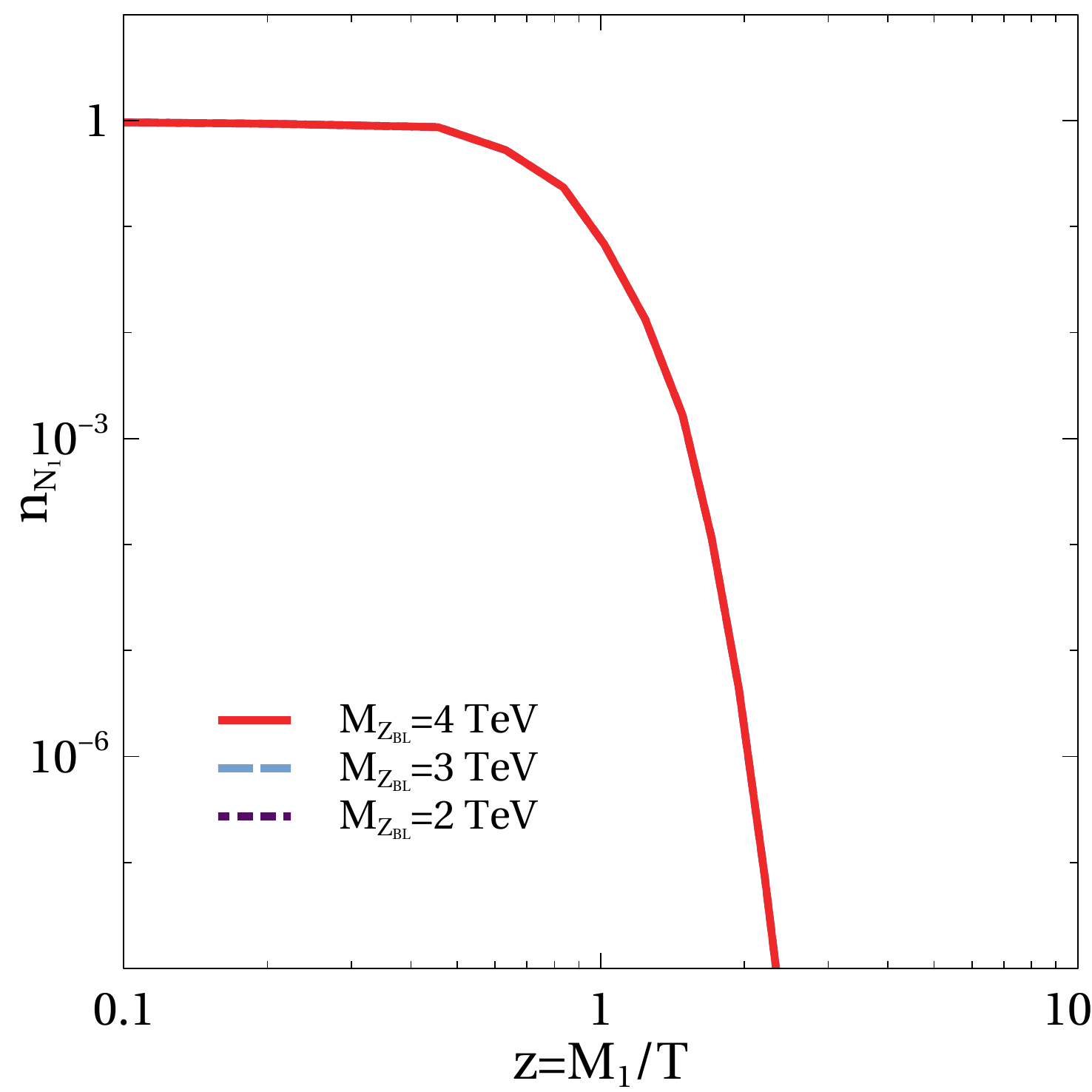}
\caption{Comoving density of B-L asymmetry (left panel) and comoving number density of $N_{R_{1}}$ (right panel) with $z=\dfrac{M_{1}}{T}$ for different $M_{Z_{BL}}$. The Yukawa couplings relevant for Leptogenesis are taken to be $(Y_{\eta})_{\alpha 1}=10^{-5}(1+i)$ and $(Y_{\eta})_{\alpha 2}=10^{-1}(1.1-i)$. The other important parameters used are $M_{1}=45$ TeV, $M_{2}=450$ TeV, $m_{\eta}=5$ TeV and $g_{BL}=10^{-3}$. The horizontal line in left panel plot denotes the required $B-L$ asymmetry to generate for observed baryon asymmetry (Planck 2018) after sphaleron transition.}
\label{leptofig2}
\end{center}
\end{figure}

\begin{figure}[h]
\begin{center}
\includegraphics[scale=.477]{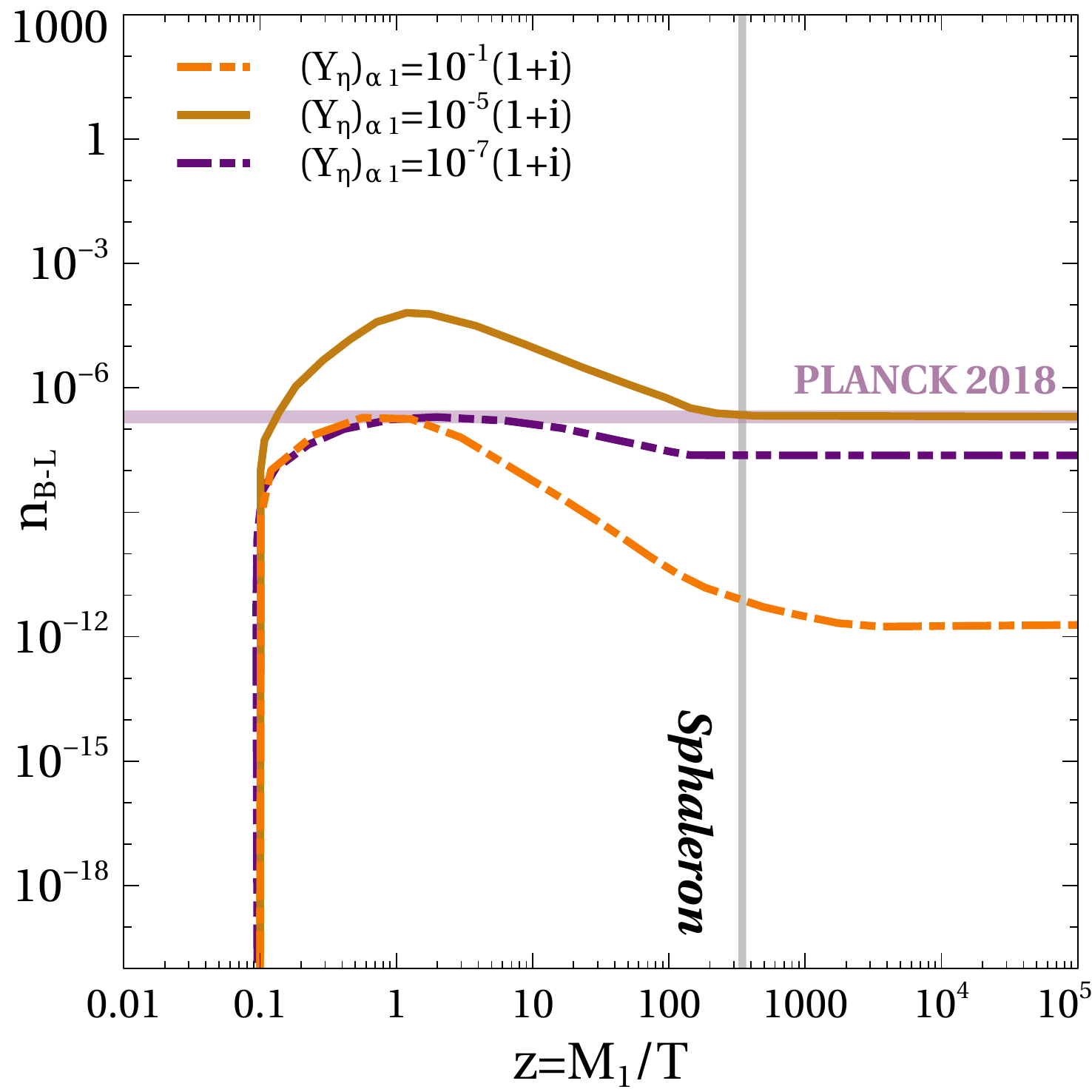}
\includegraphics[scale=.35]{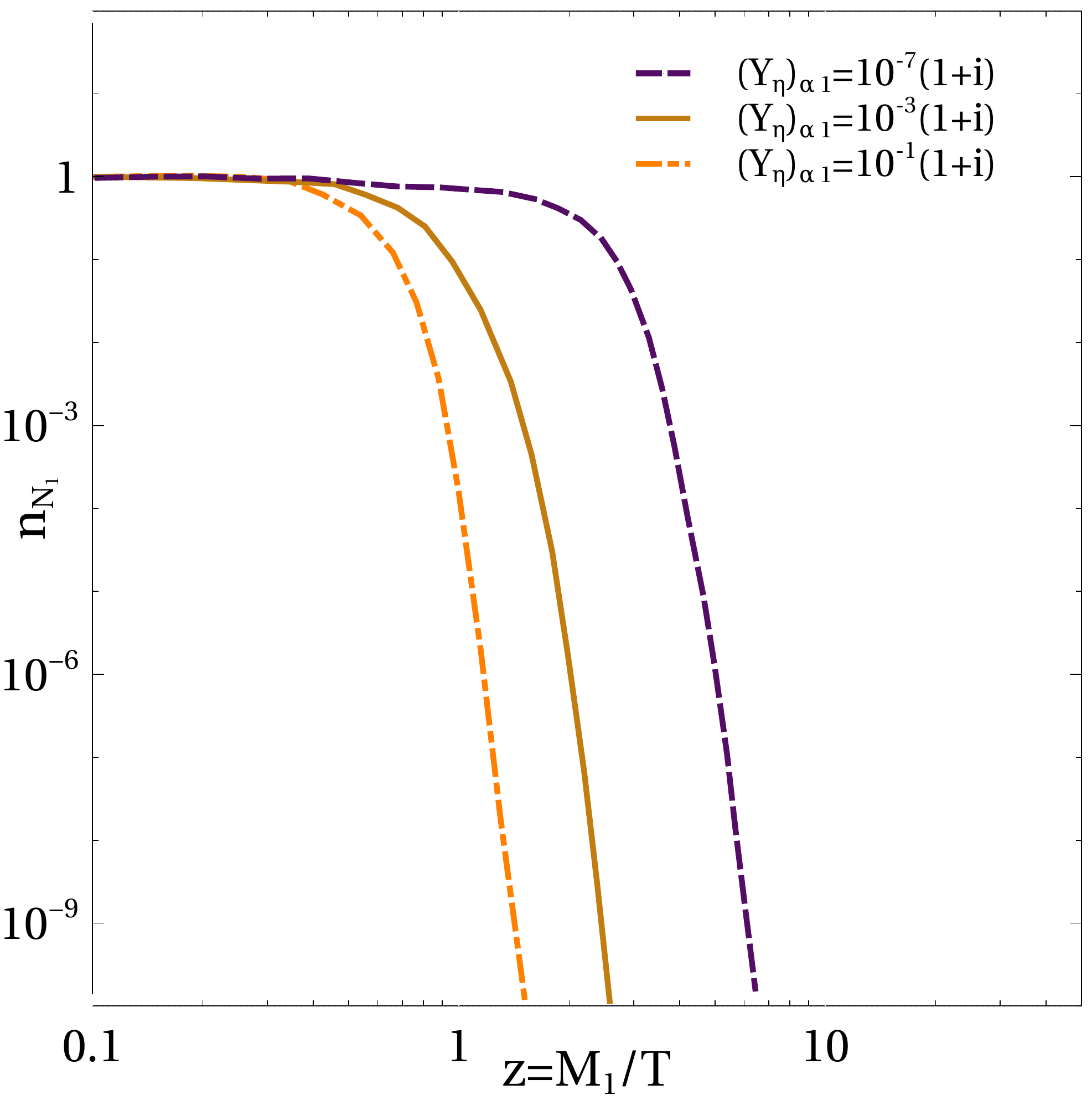}
\caption{Comoving density of B-L asymmetry (left panel) and comoving number density of $N_{R_{1}}$ (right panel) with $z=\dfrac{M_{1}}{T}$ for different $(Y_{\eta})_{\alpha 1}$. The $B-L$ gauge coupling relevant for Leptogenesis are taken to be $g_{BL}=10^{-3}$ and $(Y_{\eta})_{\alpha 2}=10^{-1}(1.1-i)$. The other important parameters used are $M_{1}=45$ TeV, $M_{2}=450$ TeV, $m_{\eta}=5$ TeV and $M_{Z_{B-L}}=4$ TeV. The horizontal line in left panel plot denotes the required $B-L$ asymmetry to generate for observed baryon asymmetry (Planck 2018) after sphaleron transition.}
\label{leptofig4}
\end{center}
\end{figure}

We first show the CP asymmetry parameter $\epsilon_1$ and the decay parameter $K_{N_{1}}=\Gamma_{1}/H(z=1)$ with heavy Majorana neutrino mass $M_1$ for different benchmark values Yukawa couplings. We consider flavour universal Yukawa couplings of $N_i$ for simplicity. From the left panel plot of figure \ref{epsilon_kappa} we can see that with a fixed value of the $N_{2}$ Yukawa coupling $(Y_{\eta})_{\alpha 2}$ the decay parameter $K_{N_{1}}$ increases with the increase in $N_{1}$ Yukawa coupling as expected whereas the CP asymmetry parameter $\epsilon_1$ remains constant. The CP asymmetry parameter $\epsilon_1$ does not change with the change in $(Y_{\eta})_{\alpha 1}$ as $(Y_{\eta})_{\alpha 2}$ decides the imaginary part of Yukawa coupling products appearing in CP asymmetry $\epsilon_1$, as described the Appendix \ref{asymmetry}. In the right panel plot of figure \ref{epsilon_kappa} one can see that with the increase in $(Y_{\eta})_{\alpha 2}$ the CP asymmetry parameter $\epsilon_1$ increases while $K_{N_{1}}$ remains same due to the fixed Yukawa coupling $(Y_{\eta})_{\alpha 1}$. The increase in CP asymmetry with the increase in $(Y_{\eta})_{\alpha 2}$ can be understood from the description of Appendix \ref{asymmetry}. The constancy of the decay parameter with respect to $(Y_{\eta})_{\alpha 2}$ is trivially understood as $N_{1}$ decay width depends on $(Y_{\eta})_{\alpha 1}$ only.From this figure \ref{epsilon_kappa} we can understand that for a fixed value of $(Y_{\eta})_{\alpha 1}$ one need relatively large value of the $(Y_{\eta})_{\alpha 2}$ to have enough CP asymmetry $\epsilon_1$. In both the plots, the decay parameter $K_{N_{1}}$ remains very large for some part of the parameter space. Such large values will lead to wash-out of the $B-L$ asymmetry via inverse decay. We mark the points as stars which correspond to the combination of Yukawa as well as $M_1$ that can give rise to correct final asymmetry. Clearly, both these points correspond to small $K_{N_{1}}$ which can be ensured for smaller Yukawa coupling $(Y_{\eta})_{\alpha 1}$ and larger $M_1$. The washout effects are also visible in the evolution plots \ref{leptofig1}, \ref{leptofig2}, \ref{leptofig4}. It should be noted that in a strong washout region ($K_{N_{1}}>1$), the generated asymmetry can be made large by choosing large enough $(Y_{\eta})_{\alpha 2}$ (within perturbative limit) such that it gives the correct asymmetry after the washout process become negligible. Since both the Yukawa couplings in our model are completely free from the light neutrino sector we can tune them accordingly to give the observed asymmetry. This is in sharp contrast with type I seesaw leptogenesis where Yukawa couplings are constrained by light neutrino mass data for a fixed scale of leptogenesis, leading to the Davidson Ibarra bound $M_1 \geq 10^9$ GeV \cite{Davidson:2002qv}. Note that increase in Yukawa coupling also increases the inverse decay, decreasing the asymmetry generated, the details of which we discuss below.

\begin{figure}[h]
\begin{center}
\includegraphics[scale=.5]{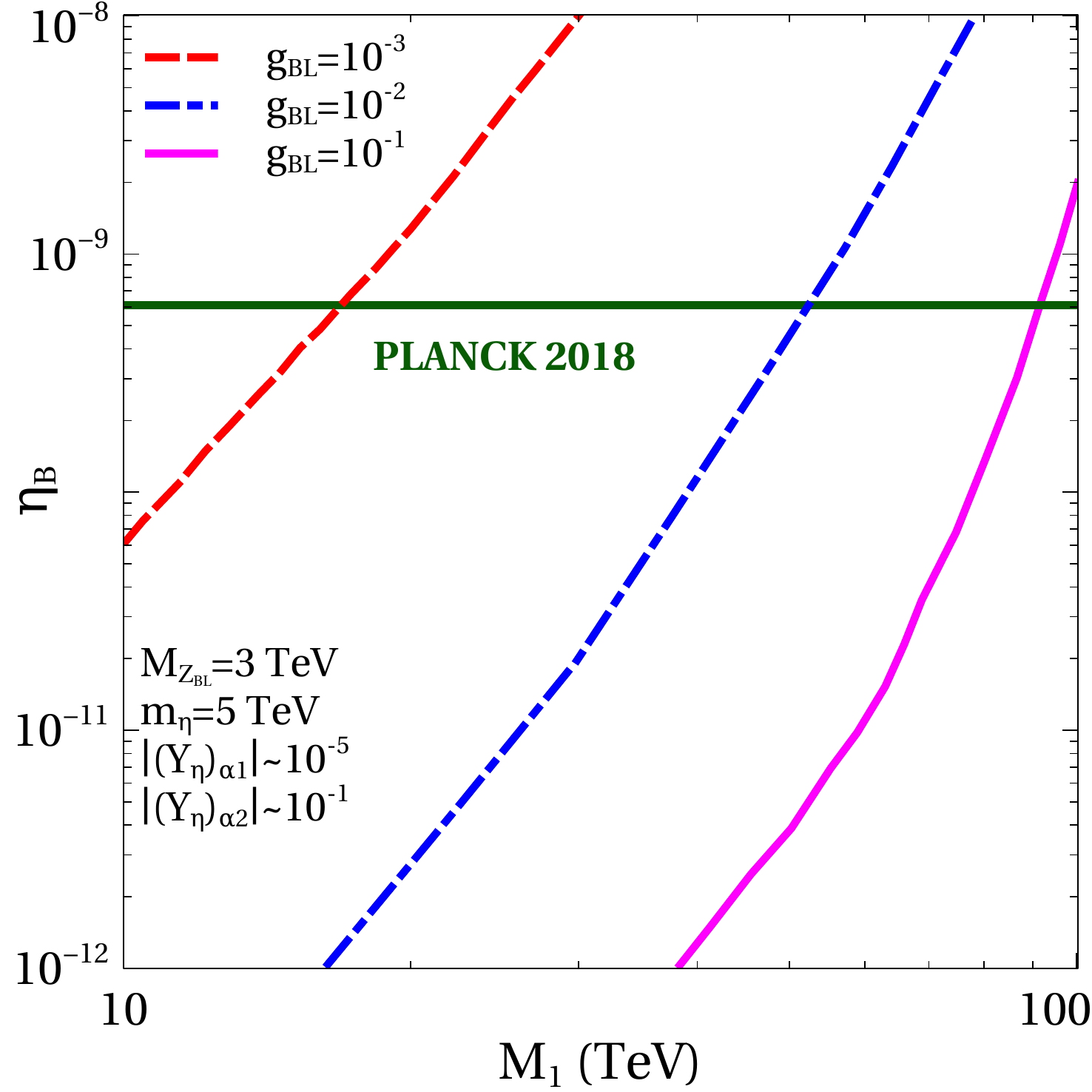}
\includegraphics[scale=.5]{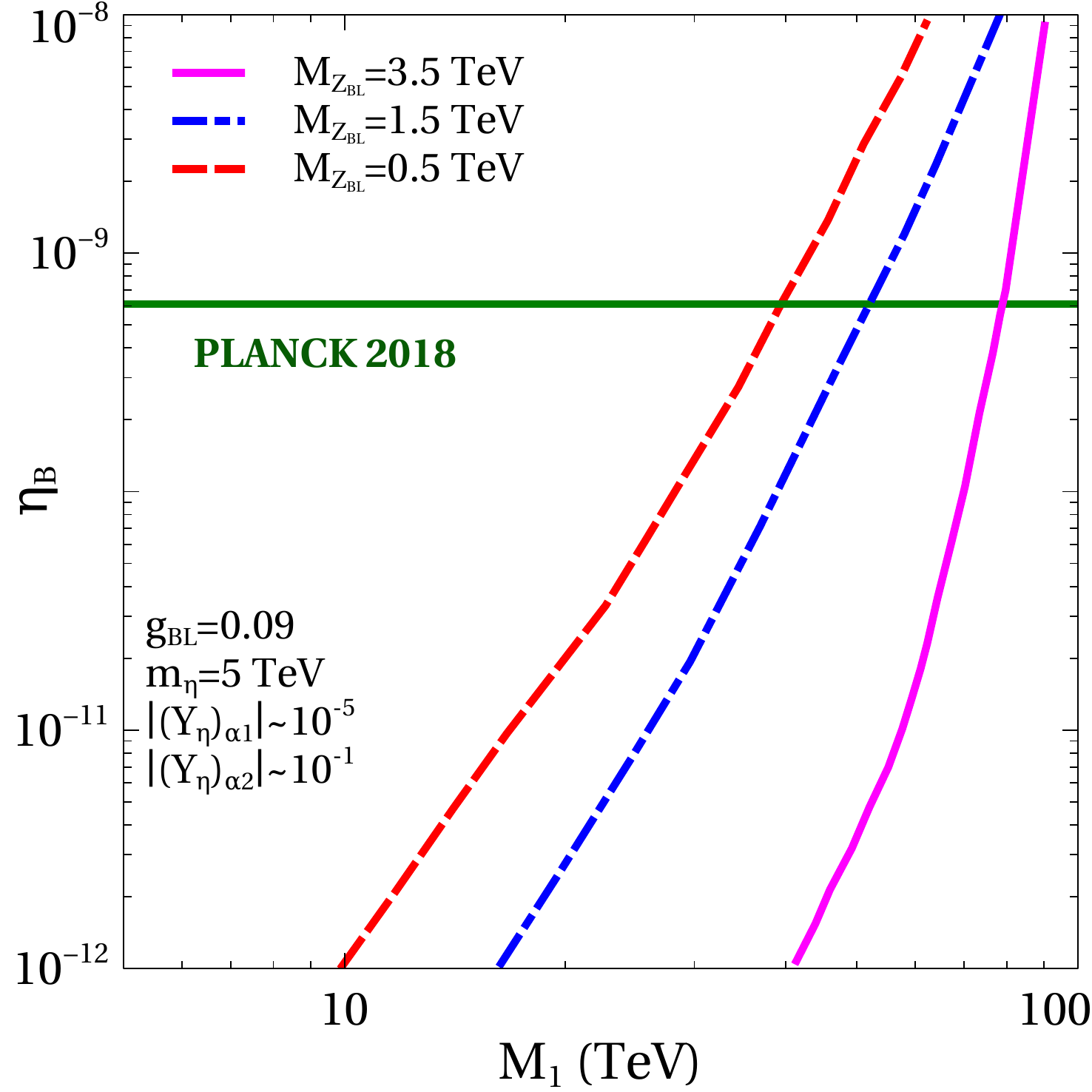}
\caption{Baryon to photon ration with mass of $N_{1}$ for different benchmark values of $g_{BL}$ (left panel) and $M_{Z_{BL}}$ (right panel). The Yukawas taken are $(Y_{\eta})_{ \alpha 1}=10^{-5}(1+i)$ and $(Y_{\eta})_{\alpha 2}=3\times10^{-1}(1.1-i)$. }
\label{leptofig3}
\end{center}
\end{figure}

In left panel plot of figure \ref{leptofig1}, we show the evolution of lepton asymmetry for different values of $B-L$ gauge couplings while fixing other parameters at benchmark values. Note that, for the numerical analysis we consider $n_1=-3/2$. The right panel plot of the same figure shows the corresponding evolution of $N_{R_1}$ number density $n_{N_1}$. The evolution of comoving number densities of $N_{1}$ for different values of $g_{BL}$ do not show significant differences as they are primarily governed by their decay. However, presence of additional gauge interactions can lead to significant changes in $N_1$ number density as they tend to keep $N_1$ in equilibrium for longer duration. We show these details in Appendix \ref{washouts}. The effect of the $N_{1}$ annihilations is visible in the evolution plots of the comoving number density of $B-L$ asymmetry in figure \ref{leptofig1} and figure \ref{leptofig2}. With the increase in $g_{BL}$ and $M_{Z_{BL}}$ the $N_{1}$ annihilation cross section increases which brings the $N_{1}$ number density closer to its equilibrium number density and therefore the $B-L$ asymmetry decreases. We observe similar behavior of the asymmetry with $g_{BL}$ and $M_{Z_{BL}}$ because of the washout process $l Z_{BL} \longrightarrow N_{1}\eta$. The details of the relevant washout is shown in Appendix \ref{washouts}. However, we noticed that in the determination of the $B-L$ asymmetry the $N_{1}$ annihilation plays more dominant role than the washout $l Z_{BL} \longrightarrow N_{1}\eta$. In the left panel plot of figure \ref{leptofig1}, the horizontal line shows the required final comoving lepton asymmetry at the epochs of sphaleron transition which gets converted into the observed baryon asymmetry. The final $B-L$ asymmetry $n_{B-L}^f$ just before electroweak sphaleron freeze-out is converted into the observed baryon to photon ratio by the standard formula 
\begin{align}
\eta_B \ = \ \frac{3}{4}\frac{g_*^{0}}{g_*}a_{\rm sph}n_{B-L}^f \ \simeq \ 9.2\times 10^{-3}\: n_{B-L}^f \, ,
\label{eq:etaB}
\end{align}
where $a_{\rm sph}=\frac{8}{23}$ is the sphaleron conversion factor (taking into account two Higgs doublets). The effective relativistic degrees of freedom is taken to be $g_*=116$, slightly higher than that of the SM at such temperatures as we are including the contribution of the inert Higgs doublet and $\nu_R$'s too. In the above expression $g_*^0=\frac{43}{11}$ is the effective relativistic degrees of freedom at the recombination epoch. Using observed $\eta_B$ from equation \eqref{etaBobs}, the required final baryon asymmetry can be found as $n_{B-L}^f \approx 7 \times 10^{-8}$ corresponding to the horizontal line labelled as Planck 2018 in figure \ref{leptofig1} (left panel).

In figure \ref{leptofig4} the evolution of comoving number density of $B-L$ and $N_{1}$ are shown for different values of $(Y_{\eta})_{\alpha 1}$ keeping the other parameters fixed. One can see a two-way behavior of the asymmetry with $(Y_{\eta})_{\alpha 1}$ in the left panel plot of figure \ref{leptofig4}. With the decrease in the Yukawa coupling $(Y_{\eta})_{\alpha 1}$ the asymmetry increases, it is mainly because with the decrease in $(Y_{\eta})_{\alpha 1}$ the scattering washouts as well as the inverse decay rate decrease significantly. However, beyond a certain small value of $(Y_{\eta})_{\alpha 1}$ the asymmetry decreases with the decrease in $(Y_{\eta})_{\alpha 1}$. The is because even though the washout rates decrease with a decrease in the Yukawa couplings, the generation of the asymmetry itself decreases because of the decrease in the decay width of $N_{1}$. Thus, the interplay of production and washout dictate the strength of required Yukawa couplings for a given scale of leptogenesis. Unlike the negligible dependence of $N_1$ number density on $B-L$ sector parameters discussed above, variation in Yukawa coupling leads to noticeable changes in $N_1$ abundance, as can be seen from the right panel plot of figure \ref{leptofig4}. This is due to strong dependence of $N_1$ lifetime on the Yukawa coupling. Larger the Yukawa coupling, quicker is the fall of of $N_1$ abundance, as expected.

In figure \ref{leptofig3}, we show the final baryon asymmetry against the scale of leptogenesis namely, the mass of $N_{1}$. As seen from this figure, scale of leptogenesis can be tens of TeV depending upon the values of $g_{BL}$ as well as $M_{Z_{BL}}$. We also check the evolution of lepton asymmetry for superheavy $Z_{BL}$ (much above the scale of leptogenesis $M_1$) and find that variation in $M_{Z_{BL}}$ in such a case does not lead to any noticeable change in $n_{B-L}$ evolution, as expected.


\begin{figure}
\begin{center}
\includegraphics[scale=.6]{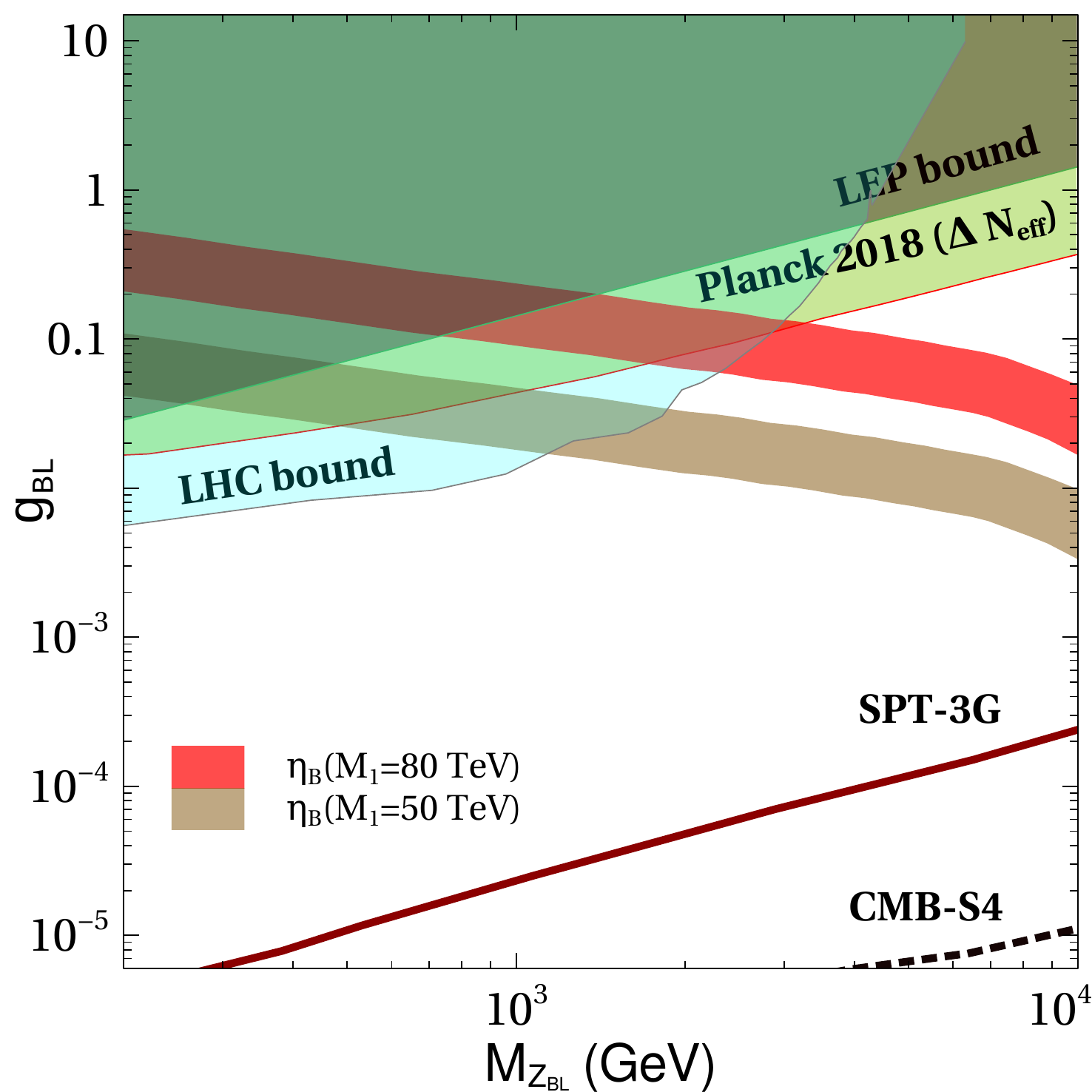}
\caption{Allowed parameter space in $g_{BL}-M_{Z_{BL}}$ plane. The red and brown coloured bands correspond to regions of successful leptogenesis for two different scales. Bounds from LEP, LHC and future sensitivities of CMB experiments are shown. The Yukawa couplings chosen for this scan are $(Y_\eta)_{\alpha 1}=10^{-5}(1+i)$ and $(Y_\eta)_{\alpha 2}=3\times 10^{-1}(1.1-i)$.}
\label{summaryfig}
\end{center}
\end{figure}



%

\subsection{Relativistic degrees of freedom $N_{\rm eff}$}
The Dirac nature of light neutrinos introduces additional relativistic degrees of freedom which can be thermalised in the early universe by virtue of $B-L$ gauge interactions. Such additional light degrees of freedom can be probed by precise measurements of the CMB anisotropies. Recent 2018 data from the CMB measurement by the Planck satellite \cite{Aghanim:2018eyx} suggests that the effective degrees of freedom for neutrinos as 
\begin{eqnarray}
{\rm
N_{eff}= 2.99^{+0.34}_{-0.33}
}
\label{Neff}
\end{eqnarray}
at $2\sigma$ or $95\%$ CL including baryon acoustic oscillation (BAO) data. At $1\sigma$ CL it becomes more stringent to $N_{\rm eff} = 2.99 \pm 0.17$. Both these bounds are consistent with the standard model (SM) prediction $N^{\rm SM}_{\rm eff}=3.045$ \cite{Mangano:2005cc, Grohs:2015tfy, deSalas:2016ztq}. Upcoming CMB Stage IV (CMB-S4) experiments are expected to put much more stringent bounds than Planck due to their potential of probing all the way down to $\Delta N_{\rm eff}=N_{\rm eff}-N^{\rm SM}_{\rm eff} = 0.06$ \cite{Abazajian:2019eic}. For some recent studies on light Dirac neutrinos and enhanced $ \Delta N_{\rm eff}$ in different contexts, please see \cite{Abazajian:2019oqj, FileviezPerez:2019cyn, Nanda:2019nqy, Han:2020oet, Luo:2020sho, Borah:2020boy, Adshead:2020ekg, Luo:2020fdt}.

Effective number of relativistic degrees of freedom is defined as 
$$ N_{\rm eff} \equiv \frac{8}{7} \left( \frac{11}{4} \right)^{4/3} \left( \frac{\rho_{\rm rad} -\rho_{\gamma}}{\rho_{\gamma}} \right) $$
where $\rho_{\rm rad}=\rho_{\gamma} + \rho_{\nu}$ is the net radiation content of the universe. As mentioned earlier, the SM prediction is $N^{\rm SM}_{\rm eff}=3.045$ \cite{Mangano:2005cc, Grohs:2015tfy, deSalas:2016ztq}. In our model, $\Delta N_{\rm eff}$ can be estimated by finding the decoupling temperature $T_{\nu_R}$ of right handed neutrinos ($\nu_R$) using 
\begin{eqnarray}
\Gamma (T_{\nu_R}^{\rm d}) = H (T_{\nu_R}^{\rm d})
\label{eq7}
\end{eqnarray}
where $\Gamma(T)$ is the interaction rate and $H(T)$ is the expansion rate of the universe. It should be noted that the decoupling is never instantaneous in reality and non-instantaneous decoupling can also lead to spectral distortions of neutrino distributions. However such spectral distortions have been found to be very small $~0.01\%$ (see \cite{Luo:2020sho}, for example). Also, the decoupling temperature calculated using instantaneous decoupling approximation above remains in qualitative agreement with full treatment incorporating relevant Boltzmann equations \cite{Luo:2020sho}. Therefore, we stick to this simplistic approach in our work here.

We show the constraints on model parameters $g_{BL}-M_{Z_{BL}}$ from Planck $2\sigma$ bound on $\Delta N_{\rm eff}$ in figure \ref{summaryfig}. The same plot also shows the LEP II limit $M_{Z_{BL}}/g_{BL} \geq 7$ TeV \cite{Carena:2004xs, Cacciapaglia:2006pk}. The LHC bound is implemented by considering the 13 TeV limit from the ATLAS experiment \cite{Aaboud:2017buh, Aad:2019fac} and CMS experiment \cite{Sirunyan:2018exx}. Clearly, the Planck $2\sigma$ bound on $\Delta N_{\rm eff}$ remains stronger than the LEP II as well as the LHC bounds for $Z_{BL}$ mass heavier than around 3 TeV. We finally show the parameter space giving rise to successful leptogenesis for two different masses of $N_{1}$ while keeping $N_2$ mass ten times higher. Clearly, lower the scale of leptogenesis, lower should be the gauge coupling $g_{BL}$ in order to keep the washout processes suppressed. Additionally, for fixed $g_{BL}$ the scale of leptogenesis gets pushed up for larger values of $M_{Z_{BL}}$ as heavier $Z_{BL}$ leads to increase in $l Z_{BL}\longrightarrow N_{1}\eta$ washout process noted earlier.

While it is possible to obtain successful leptogenesis at a scale as low as a TeV, the required gauge coupling $g_{BL}$ for such scenarios will be insufficient to generate correct WIMP DM phenomenology as we discuss below. Interestingly, the next generation CMB experiments like CMB-S4 \cite{Abazajian:2016yjj},
SPT-3G \cite{Benson:2014qhw} (whose sensitivities are shown as dashed and solid lines respectively) will be able to probe the entire parameter space consistent with successful Dirac leptogenesis.

\section{Anomaly Free $B-L$ Model}
\label{sec2}
After discussing the interesting features related to Dirac leptogenesis and observable $\Delta N_{\rm eff}$ in previous section, we now move onto discussing the complete model, which is required to be anomaly free. While the SM fermion content with gauged B-L symmetry leads to triangle anomalies, including three right handed neutrinos make the model anomaly free. Therefore, including additional heavy fermions for realising successful Dirac leptogenesis introduces new anomalies as well. In this section, we briefly discuss one such possibility where the anomalies introduced by heavy chiral fermions required for Dirac leptogenesis as well as dark matter cancel among themselves leading to an anomaly free scenario.

Sticking to a minimal setup, we consider two heavy Majorana fermions $N_R$, sufficient to produce non-vanishing CP asymmetry. Since we do not want to violate lepton number by two units, therefore, we assign $B-L$ charge $-3/2$ each to $N_R$, as earlier. This is just a choice for numerical calculations, any other $B-L$ gauge charge for $N_R$ will not change our results drastically, as long as we can ensure Dirac nature of light neutrinos. In order to make it couple to usual leptons, we introduce another Higgs doublet $\eta$ having $B-L$ charge $-1/2$ and its neutral component does not acquire any VEV. A singlet scalar $\phi_1$ having $B-L$ charge 3 is introduced in order to give mass to $N_R$ after spontaneous gauged $B-L$ symmetry breaking.  However, the introduction of these two heavy Majorana fermions again gives rise to triangle anomalies as
\begin{align}
\mathcal{A}_1 \left[ U(1)^3_{B-L} \right] =54/8  \nonumber \\
\mathcal{A}_2 \left[(\text{gravity})^2 \times U(1)_{B-L} \right] =3.
\end{align}

These anomalies can be cancelled after introducing four chiral fermions $\chi_L, \chi_R, \psi_L, \psi_R$ having $B-L$ charges $9/4, 5/4, -15/8, 17/8$ respectively. This can be seen as 
\begin{align}
\mathcal{A}_1 \left[ U(1)^3_{B-L} \right] = \left( \frac{9}{4} \right)^3 +\left( -\frac{5}{4} \right)^3+\left( -\frac{15}{8} \right)^3+\left( -\frac{17}{8} \right)^3=-54/8  \nonumber \\
\mathcal{A}_2 \left[(\text{gravity})^2 \times U(1)_{B-L} \right] =\left( \frac{9}{4} \right) +\left( -\frac{5}{4} \right)+\left( -\frac{15}{8} \right)+\left( -\frac{17}{8} \right) = -3
\end{align}
While this solution is not unique, we stick to this minimal solution which leads to the desired phenomenology without requiring arbitrarily large $B-L$ charges or more chiral fermions than the above-mentioned ones. One can also pursue non-minimal scenarios which can explain tiny Dirac neutrino masses dynamically, as done in earlier works \cite{Babu:1988yq, Peltoniemi:1992ss, Chulia:2016ngi, Aranda:2013gga, Chen:2015jta, Ma:2015mjd, Reig:2016ewy, Wang:2016lve, Wang:2017mcy, Wang:2006jy, Gabriel:2006ns, Davidson:2009ha, Davidson:2010sf, Bonilla:2016zef, Farzan:2012sa, Bonilla:2016diq, Ma:2016mwh, Ma:2017kgb, Borah:2016lrl, Borah:2016zbd, Borah:2016hqn, Borah:2017leo, CentellesChulia:2017koy, Bonilla:2017ekt, Memenga:2013vc, Borah:2017dmk, CentellesChulia:2018gwr, CentellesChulia:2018bkz, Han:2018zcn, Borah:2018gjk, Borah:2018nvu, CentellesChulia:2019xky,Jana:2019mgj, Borah:2019bdi, Dasgupta:2019rmf, Correia:2019vbn, Ma:2019byo, Ma:2019iwj, Baek:2019wdn, Saad:2019bqf, Jana:2019mez, Nanda:2019nqy}. We, however, stick to the minimal way of generating Dirac neutrino masses from SM Higgs at the cost of fine-tuned Dirac Yukawa coupling.

The fermion and scalar content of the model are shown in table \ref{tab:data1} and \ref{tab:data2} respectively. The necessity of the individual scalar fields will be discussed later. 
\begin{table}
\begin{center}
\begin{tabular}{|c|c|}
\hline
Particles & $SU(3)_c \times SU(2)_L \times U(1)_Y \times U(1)_{B-L} $   \\
\hline
$q_L=\begin{pmatrix}u_{L}\\
d_{L}\end{pmatrix}$ & $(3, 2, \frac{1}{6}, \frac{1}{3})$  \\
$u_R$ & $(3, 1, \frac{2}{3}, \frac{1}{3})$  \\
$d_R$ & $(3, 1, -\frac{1}{3}, \frac{1}{3})$  \\

$\ell_L=\begin{pmatrix}\nu_{L}\\
e_{L}\end{pmatrix}$ & $(1, 2, -\frac{1}{2}, -1)$  \\
$e_R$ & $(1, 1, -1, -1)$ \\
$\nu_R$ & $ (1, 1, 0, -1)$ \\
\hline
$N_{R1}$ & $ (1, 1, 0, -3/2)$ \\
$N_{R2}$ & $ (1, 1, 0, -3/2)$ \\
\hline
$\chi_L$ & $(1, 1, 0, \frac{9}{4})$ \\
$\chi_R$ & $(1, 1, 0, \frac{5}{4})$ \\
$\psi_L$ & $(1, 1, 0, -\frac{15}{8})$ \\
$\psi_R$ & $(1, 1, 0, \frac{17}{8})$ \\
\hline

\end{tabular}
\end{center}
\caption{Fermion Content of the Model}
\label{tab:data1}
\end{table}
\begin{table}
\begin{center}
\begin{tabular}{|c|c|}
\hline
Particles & $SU(3)_c \times SU(2)_L \times U(1)_Y \times U(1)_{B-L} $   \\
\hline
$H=\begin{pmatrix}H^+\\
H^0\end{pmatrix}$ & $(1,2,\frac{1}{2},0)$  \\
$\eta=\begin{pmatrix}\eta^+\\
\eta^0\end{pmatrix}$ & $(1,2,\frac{1}{2}, -\frac{1}{2})$  \\
\hline
$\phi_1$ & $(1, 1, 0, 3)$ \\
$\phi_2$ & $(1, 1, 0, 4)$ \\
$\phi_3$ & $(1, 1, 0, 1)$ \\
\hline
\end{tabular}
\end{center}
\caption{Scalar content of the Minimal Model}
\label{tab:data2}
\end{table}

The relevant Yukawa Lagrangian is 
\begin{align}
\mathcal{L}_Y \supset Y_D \bar{L} \tilde{H} \nu_R + Y_{\eta} \bar{L} \tilde{\eta} N_R + Y_N \phi_1 N_R N_R+ Y_{\chi} \bar{\chi_L} \chi_R \phi_3 + Y_{\psi} \bar{\psi_L} \psi_R \phi^{\dagger}_2+{\rm h.c.}
\end{align}
Here also, Dirac leptogenesis may occur through out-of-equilibrium decay of $N_R$'s. There are several DM candidates here in terms of $\chi, \psi, \eta$.

The
gauge invariant scalar interactions described by $\mathcal{L}_{scalar}$ can be written as  
\begin{align}
\mathcal{L}_{scalar} &= \left({D_{H}}_{\mu} H \right)^\dagger
\left({D_{H}}^{\mu} H \right) + \left({D_{\eta}}_{\mu} \eta \right)^\dagger \left({D_{\eta}}^{\mu} \eta \right) + \sum_{i=1}^3 \left({D_{\phi_i}}_{\mu} \phi_i \right)^\dagger
\left({D_{\phi_i}}^{\mu}\,\phi_i \right)- \bigg \{-\mu^2_H \lvert H \rvert^2  \nonumber \\ 
& + \lambda_H \lvert H \rvert^4 + \left( \mu^2_{\eta} \lvert \eta \rvert^2 + \lambda_{\eta} \lvert \eta \rvert^4 \right)+\sum_{i=1,2, 3} \left( -\mu^2_{\phi_i} \lvert \phi_i \rvert^2 + \lambda_{\phi_i} \lvert \phi_i \rvert^4 \right)  +\lambda_{H\eta} (\eta^{\dagger} \eta) (H^{\dagger} H)\nonumber \\
& + \lambda^{\prime}_{H\eta} (\eta^{\dagger} H) (H^{\dagger} \eta)  + \sum_{i=1,2, 3}\lambda_{H\phi_i} (\phi^{\dagger}_i \phi_i) (H^{\dagger} H) +  \sum_{i=1,2, 3} \lambda_{\eta \phi_i} (\eta^{\dagger} \eta)(\phi^{\dagger}_i \phi_i) \nonumber \\
& +\lambda_{12} (\phi^{\dagger}_1 \phi_1)(\phi^{\dagger}_2 \phi_2)+\lambda_{13} (\phi^{\dagger}_1 \phi_1)(\phi^{\dagger}_3 \phi_3)+\lambda_{23} (\phi^{\dagger}_2 \phi_2)(\phi^{\dagger}_3 \phi_3)   \nonumber \\
& + \left (\mu_{\phi} \phi_1 \phi^{\dagger}_2 \phi_3 + \text{h.c.} \right) \bigg \}
\label{scalar:lag}
\end{align}

Where $\rm{{D_{H}}^{\mu}}$, $\rm{{D_{\eta}}^{\mu}}$ and $\rm{{D_{\phi}}^{\mu}}$ denote the covariant derivatives for the scalar doublets H, $\rm{\eta}$ and scalar singlets ${\rm\phi_i}$ respectively and can be written as 

\begin{eqnarray}
{D_{H}}_{\mu}\,H &=& \left(\partial_{\mu} + i\,\dfrac{g}{2}\,\sigma_a\,W^a_{\mu}
+ i\,\dfrac{g^\prime}{2}\,B_{\mu}\right)H \,, \nonumber \\
{D_{\eta}}_{\mu}\,\eta &=& \left(\partial_{\mu} + i\,\dfrac{g}{2}\,\sigma_a\,W^a_{\mu}
+ i\,\dfrac{g^\prime}{2}\,B_{\mu} + i\,g_{BL}\,n_{\eta}
{Z_{BL}}_{\mu}\right)\eta \,, \nonumber \\
{D_{\phi}}_{\mu}\,\phi_i &=& \left(\partial_{\mu} + i\,g_{BL}\,n_{\phi_i}
{Z_{BL}}_{\mu}\right)\phi_i\,.
\end{eqnarray}
where ${\rm g_{BL}}$ is the new gauge coupling and ${\rm n_{\eta}}$ and ${\rm n_{\phi_i}}$ are the charges under ${\rm U(1)_{B-L}}$ for ${\rm \eta}$ and ${\rm \phi_i}$ respectively. After both $B-L$ and electroweak symmetries get broken by the VEVs of H and $\phi_i$s the doublet and all three
singlets are given by

\begin{eqnarray}
H=\begin{pmatrix}H^+\\
\dfrac{h^{\prime} + v + i z}{\sqrt{2}}\end{pmatrix}\,,\,\,\,\,\,\,
\eta=\begin{pmatrix}\eta^+\\
\dfrac{\eta_R^{\prime} + i \eta_I^\prime}{\sqrt{2}}\end{pmatrix}\,,\,\,\,\,\,\,
\phi_i = \dfrac{s^{\prime}_i +u_i+ A^{\prime}_i}{\sqrt{2}}\,\,\,\,(i=1,\,2, \,3)\,\,,
\label{H&phi_broken_phsae}
\end{eqnarray} 
The details of the scalar mass spectrum can be found in appendix \ref{scalar2}.


Since none of the scalar field acquiring non-zero VEV has $B-L$ charge $\pm 2$, there is no possibility of generating Majorana light neutrino masses. However, out-of-equilibrium decay $N_R \rightarrow L \eta$ can give rise to non-zero CP asymmetry in a way similar to vanilla leptogenesis, as seen from figure \ref{Fig:feyn}. However, the corresponding Yukawa couplings do not play any role in neutrino mass generation and hence are unconstrained. This allows the possibility of low scale leptogenesis that too with Dirac neutrinos. Additionally, the model also offers several dark matter candidates in terms of $\eta, \chi, \psi$. Out of them, the scalar doublet can not give rise to desired DM phenomenology due to large direct detection cross section mentioned earlier. Therefore, we keep its mass fixed at benchmark values where its relic abundance is sub-dominant.

Similar to the proposal in \cite{Heeck:2013vha}, one can also generate light Dirac neutrino mass by a neutrinophillic Higgs doublet which gets induced VEV after EWSB. An additional $Z_2$ symmetry was introduced in addition to $U(1)_{B-L}$ gauge symmetry. However, we stick to this minimal field content at the cost of fine-tuned Yukawa couplings. The conclusions reached here will not change significantly if we adopt such non-minimal scenarios. Additionally, generation of lepton asymmetry in our proposal is different from earlier works on Dirac leptogenesis. For example, in \cite{Heeck:2013vha} the CP asymmetry was generated by scalar singlet decay into right handed neutrinos $\nu_R$ through $\Delta (B-L)=4$ process. The lepton asymmetry in $\nu_R$ then gets transferred to the left sector via Yukawa interactions with the neutrinophillic Higgs doublet. This is in a way complementary to the proposal in \cite{Dick:1999je, Murayama:2002je} where an equal and opposite amount of lepton asymmetry were generated in right and left sectors (vanishing net lepton number violation) which were prevented from equilibration by virtue of tiny Dirac Yukawa couplings. On the other hand, in our proposal, leptogenesis remains very similar to vanilla leptogenesis except for the fact that the couplings involved do not play any role in neutrino mass generation and we do not have $\Delta (B-L) =2$ processes.  Additional advantage is that the model also predicts stable dark matter candidates in terms of additional chiral fermions added to cancel chiral anomalies.

Thermal dark matter in gauged $B-L$ model has been discussed by several authors, see \cite{Okada:2010wd, Basak:2013cga, Okada:2016gsh, Okada:2018ktp, Escudero:2018fwn, Nanda:2017bmi, Biswas:2019ygr, Nanda:2019nqy, Borah:2018smz, Borah:2020wyc} for example. If we consider SM singlet fermions with non-trivial $B-L$ charges to be DM candidates, the only interaction between DM and SM particles is the $B-L$ gauge interactions \footnote{Singlet scalars can also mediate DM-SM interactions due to their mixing with the SM Higgs. However, we neglect such scalar portal interactions in order to obtain maximum constraints on $B-L$ gauge sector.}. In order to calculate thermal averaged cross sections as well as relic abundance numerically, we use the package \texttt{micrOMEGAs} \cite{Belanger:2014vza}
where the necessary model information have been provided using package
\texttt{FeynRules} \cite{Alloul:2013bka}. Since DM interacts with the SM bath only via $B-L$ gauge portal interactions, relic abundance is typically satisfied only around the resonance regime $2 m_{\rm DM} = M_{Z_{BL}}$. Note that the scalar doublet $\eta$ is also stable and hence its neutral component can, in principle, be DM candidate as well. However, the neutral scalar and pseudoscalar components of $\eta$ namely $\eta_{R}^{'}, \eta_{I}^{'}$ are degenerate and hence will lead to a large DM-nucleon scattering cross section (mediated by both $Z$ and $Z_{BL}$) ruled out by direct detection experiments like XENON1T \cite{Aprile:2017iyp, Aprile:2018dbl}. The situation is similar to sneutrino DM in minimal supersymmetric standard model (MSSM) \cite{Arina:2007tm}. The only way to keep our model consistent with direct detection data is to choose $\eta$ mass and other model parameters in such a way that leads to a sub-dominant contribution to DM. While scalar doublet DM has a mass regime giving rise to under-abundant relic abundance \cite{LopezHonorez:2006gr, Borah:2017dfn}, we find that usual SM portal interactions are not sufficient to keep relic abundance of $\eta$ sub-dominant in required amount. Interestingly, it turns out that the under-abundance criteria for $\eta$ also restricts the $B-L$ gauge sector parameters. In fact, $\eta$ abundance decreases sharply, once $\eta \eta \longrightarrow Z_{BL} Z^*_{BL} \longrightarrow Z_{BL} f \bar{f}$ and $\eta \eta \longrightarrow Z_{BL} Z_{BL}$ open up. Therefore, we keep $\eta$ mass above the required threshold $(M_{Z_{BL}}/2)$ to allow at least one of these processes to contribute significantly to its thermal relic.

While the details of leptogenesis remain same as before, we incorporate additional constraints from required DM phenomenology here for the anomaly free $B-L$ model. The relevant interactions of extra chiral fermions can be written as

\begin{align}\label{eq:7}
\mathcal{L}_{DM} & =i \left[  \bar{\chi_{L}}\slashed{D}(Q_{\chi}^{L})\chi_{L}+ \bar{\chi_{R}}\slashed{D}(Q_{\chi}^{R})\chi_{R}+ \bar{\psi_{L}}\slashed{D}(Q_{\psi}^{L})\psi_{L} + \bar{\psi_{R}}\slashed{D}(Q_{\psi}^{R})\psi_{R}+   \right]-   \nonumber \\ & (f_{1}\bar{\chi_{L}}\chi_{R}\phi_{3}+f_{2}\bar{\psi_{L}}\psi_{R}\phi_{2}^{\dagger}+{\rm h.c.})
\end{align}

We now rewrite the above Lagrangian in the basis $\xi_{1}=\chi_{L}+\chi_{R}$ and $\xi_{2}=\psi_{L}+\psi_{R}$, identified as the two Dirac fermion DM candidates. In the basis of $\xi_{1}$  and $\xi_{2}$ , the above Lagrangian  can be written as

\begin{align}\label{eq:8}
\mathcal{L_{DM}} & =i\xi_{1}\slashed{\partial}\xi_{1}+i\xi_{2}\slashed{\partial}\xi_{2}-g_{BL}\left( \dfrac{9}{4}\right)\bar{\xi_{1}}\slashed{Z}_{B-L}P_{L}\xi_{1}-g_{BL}\left( \dfrac{-15}{8}\right)\bar{\xi_{2}}\slashed{Z}_{B-L}P_{L}\xi_{2} \nonumber \\ & -g_{BL}\left( \dfrac{5}{4}\right)\bar{\xi_{1}}\slashed{Z}_{B-L}P_{R}\xi_{1}-g_{BL}\left( \dfrac{17}{8}\right)\bar{\xi_{2}}\slashed{Z}_{B-L}P_{R}\xi_{2} - f_{1}\bar{\xi_{1}}P_{R}\xi_{1}\phi_{3} \nonumber \\ & -f_{1}\bar{\xi_{1}}P_{L}\xi_{1}\phi_{3}^{\dagger}-f_{2}\bar{\xi_{2}}P_{R}\xi_{2}\phi_{2}^{\dagger}-f_{2}\bar{\xi_{2}}P_{L}\xi_{2}\phi_{2}.
\end{align}

\begin{figure}
\begin{center}
\includegraphics[scale=.6]{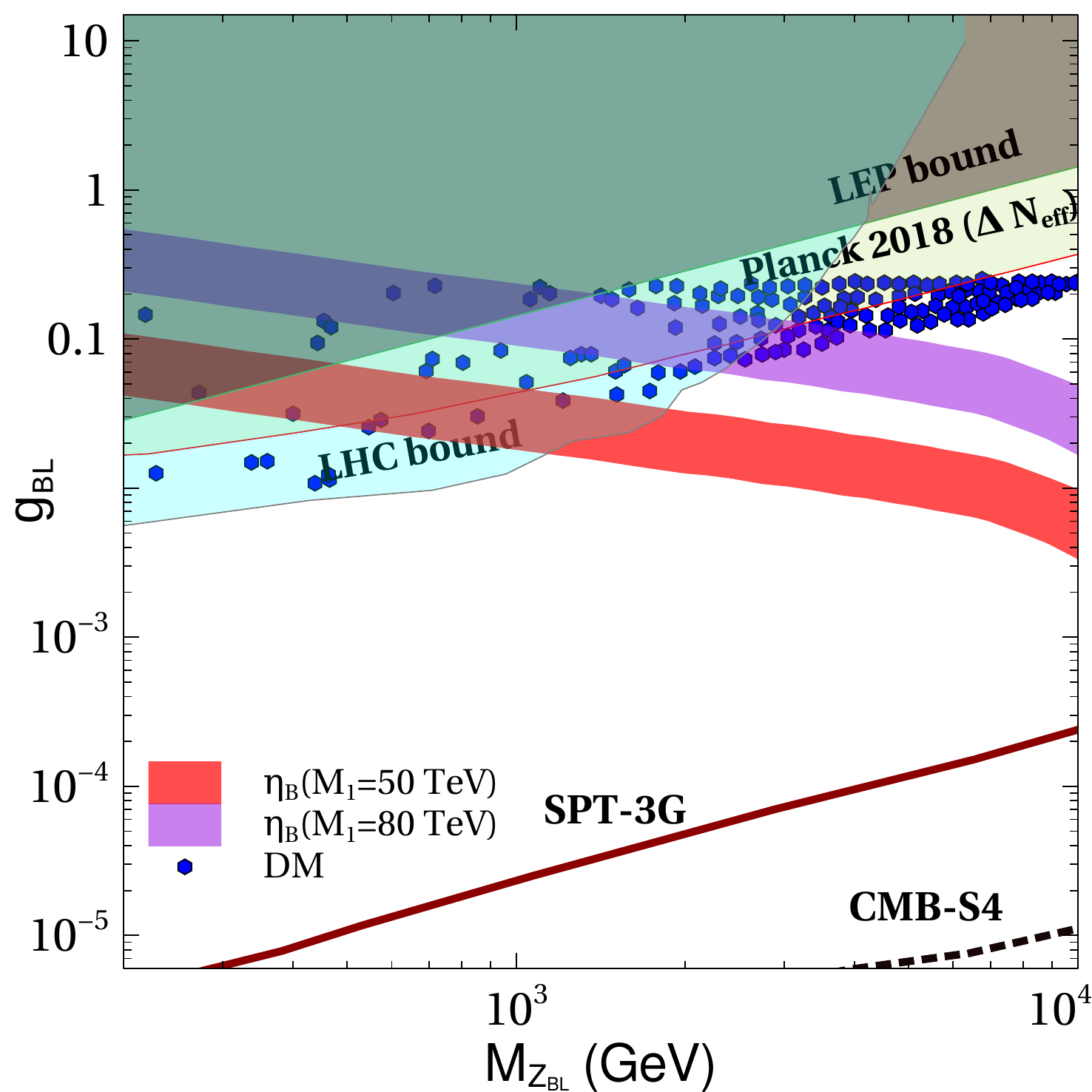}
\caption{Summary plot showing allowed parameter space in $g_{BL}-M_{Z_{BL}}$ plane. In addition to the parameter space shown already in figure \ref{summaryfig}, the allowed points from DM phenomenology are indicated by blue dots. The Yukawa couplings chosen for this scan are $(Y_\eta)_{\alpha 1}=10^{-5}(1+i)$ and $(Y_\eta)_{\alpha 2}=3\times 10^{-1}(1.1-i)$.}
\label{summaryfig2}
\end{center}
\end{figure}

Since we have two stable DM candidates i.e. $\xi_1$ and $\xi_2$ in this model comprising the dominant part of DM, the total relic abundance can be expressed as the sum of the individual candidates, ${\rm\Omega_{DM} h^2 = \Omega_{\xi_1} h^2 + \Omega_{\xi_2} h^2}$. Note that we focus particularly on $B-L$ gauge portal interactions of DM and accordingly show the parameter space satisfying total DM relic abundance in $g_{BL}-M_{Z_{BL}}$ plane of summary plot shown in figure \ref{summaryfig2}. We also found that the DM parameter space shown in figure \ref{summaryfig2} survives the direct detection bounds from XENON1T experiment. To be more quantitative, all the blue dots in figure \ref{summaryfig2} corresponds to correct total DM relic, dominantly from two fermion DM candidates and effective DM-nucleon cross section below XENON1T upper bound. The same points also correspond to sub-dominant $\eta$ (at least four to five order of magnitudes suppressed compared to observed DM relic) and hence an acceptable DM-nucleon scattering rate. Interestingly, even though $\eta$'s contribution to DM relic is negligible, its direct detection rate still lies within two to three order of magnitudes below XENON1T upper limit and should be accessible at future direct search experiments. For details of such multi-component Dirac fermion DM studies in gauge $B-L$ models, one may refer to \cite{Bernal:2018aon, Biswas:2019ygr, Nanda:2019nqy}. Here we show only the final parameter space and compare it with the ones required for other desired phenomenology like leptogenesis and $\Delta N_{\rm eff}$. 

In order to generate the summary plot of figure \ref{summaryfig2}, we randomly vary $M_{Z_{BL}}, g_{BL}$ in range $0.5 \; {\rm TeV}-10 \; {\rm TeV}$ and $0.0001-10$ respectively. Then we define $m_{\eta}=M_{Z_{BL}}-\Delta M$, where we randomly vary the mass splitting as $0.5 \; {\rm TeV} <\Delta M<0.5 M_{Z_{BL}}$ so that the minimum value of $\eta$ mass remains $M_{Z_{BL}}/2$ in order to allow its annihilation into at at least one on-shell $Z_{BL}$ discussed earlier. Note that the scalar potential of the model still allows us to choose the mass of charged component of $\eta$ to be much heavier while keeping the neutral components degenerate at $m_{\eta}$. Finally we select the points which satisfy the Planck 2018 relic bound for the total DM relic with each DM component satisfying direct detection bounds. Similarly, for the same variation of parameters, we check the parameter space giving rise to successful leptogenesis for two different scale of leptogenesis.

In the summary plot of figure \ref{summaryfig2}, we show the parameter space for successful leptogenesis with scale $M_1=50$ TeV, 80 TeV respectively. Also, for simplicity, we have considered the singlet scalars to be much heavier than DM as well as $Z_{BL}$ so that the scalar portal interactions are sub-dominant. Including scalar portal interactions will widen up the DM parameter space further due to less dependence on gauge portal annihilations. Thus, as expected, one can achieve successful leptogenesis even with hierarchical heavy neutrinos as light as a few tens of TeV. This is particularly due to the fact that the Yukawa couplings as well as heavy fermion masses which dictate the dynamics of leptogenesis are decoupled from light neutrino mass generation in our model. In fact, the scale of leptogenesis can be as low as a TeV also, with smaller values of $g_{BL}$. However, this will not be sufficient to keep the abundance of $\eta$ sub-dominant in order to keep its direct detection rate within limits.

We also show the bounds from collider experiments as well as Planck bound on $\Delta N_{\rm eff}$ at $2\sigma$ in the plot of figure \ref{summaryfig2}. Clearly, a large part of the parameter space satisfying correct DM phenomenology and leptogenesis criteria is disfavoured by these bounds. In fact, for $M_1=50$ TeV, the common parameter space satisfying DM and leptogenesis criteria are ruled out while a small part of the parameter space for $M_1=80$ TeV still remain allowed. On the other hand, the currently allowed parameter space can be completely probed by future measurements of $\Delta N_{\rm eff}$ at upcoming CMB experiments, keeping the model testable at near future.

\section{Conclusion}
\label{sec3}
We have proposed a minimal scenario to achieve successful Dirac leptogenests at a scale of few tens of TeV along with dark matter by considering a gauged $B-L$ extension of the standard model. While light Dirac neutrino mass is assumed to arise from SM Higgs coupling for simplicity, additional heavy Majorana fermions are introduced in such a way that lead to $\Delta (B-L) \neq 2$ processes, required to prevent Majorana mass contribution to light neutrinos. These heavy neutrinos can couple to SM lepton doublets via an additional Higgs doublet and hence can lead to the generation of lepton asymmetry through standard out-of-equilibrium decays. Since the relevant Yukawa couplings involved in such decays remain decoupled from light neutrino mass generation, they can be tuned freely so as to realise successful Dirac leptogenesis at low scale. Since these heavy Majorana fermions have gauged $B-L$ interactions, the corresponding gauge washout effects play non-trivial role thereby giving constraints on $g_{BL}-M_{Z_{BL}}$ parameter space from successful leptogenesis criteria. Additionally, the Dirac nature of light neutrinos lead to additional relativistic degrees of freedom which can be thermalised due to gauged $B-L$ interactions and hence can be constrained further from Planck 2018 bounds on such additional thermalised relativistic degrees of freedom $\Delta N_{\rm eff}$.

While the additional scalar doublet $\eta$ can be a stable DM candidate by itself, its degenerate neutral components give rise to a large direct detection rate, similar to sneutrino DM in MSSM. This requires sub-dominant contribution of $\eta$ to DM relic and we show that it can be achieved only when its annihilations into $Z_{BL}$ becomes kinematically allowed, further constraining $g_{BL}-M_{Z_{BL}}$ parameter space from the requirement of its under-abundance. On the other hand, additional chiral fermions are needed to make the model anomaly free. We show one possible way of cancelling these anomalies with the inclusion of four chiral fermions with fractional $B-L$ charges. With appropriate choice of singlet scalars, these fermions give rise to two Dirac fermions, eligible for being DM candidates. We show that the criteria of correct DM relic from fermion DM along with sub-dominant scalar doublet while being in agreement with direct detection data constrains the $g_{BL}-M_{Z_{BL}}$ parameter space significantly. For two different scales of leptogenesis we considered, only the higher one namely, $M_1=80$ TeV gives rise to some common parameter space consistent with all criteria and experimental bounds. Future LHC runs as well as CMB measurements will be able to probe the entire parameter space consistent with successful Dirac leptogenesis and thermal dark matter.

\acknowledgements
DB acknowledges the support from Early Career Research Award from DST-SERB, Government of India (reference number: ECR/2017/001873). DM would like to thank Dibyendu Nanda for useful discussions during the course of this project.

\appendix

\section{Scalar mass matrix diagonalisation}
\label{scalar2}
After putting equation \eqref{H&phi_broken_phsae} in equation \eqref{scalar:lag}  we have found out the $5\times5$ mixing matrix for the real scalar fields in the basis $\dfrac{1}{\sqrt{2}}$ $\begin{pmatrix} h^{'} & \eta_{R}^{'} & S_{1}^{'} & S_{2}^{'} & S_{3}^{'} 
\end{pmatrix}$ which has the following form
\begin{eqnarray}
\begin{pmatrix}
m_{hh}^{'} & 0 & m_{hS_{1}}^{} & m_{hS_{2}}^{'} & m_{hS_{3}}^{'} \\
0 & m_{\eta_{R}\eta_{R}}^{'} & 0 & 0 & 0 \\
m'_{S_{1}h} & 0 & m'_{S_{1}S_{1}} & m'_{S_{1}S_{2}} & m'_{S_{1}S_{3}} \\
m'_{S_{2}h} & 0 & m'_{S_{2}S_{1}} & m'_{S_{2}S_{2}} & m'_{S_{2}S_{3}} \\
m'_{S_{3}h} & 0 & m'_{S_{3}S_{1}} & m'_{S_{3}S_{2}} & m'_{S_{3}S_{3}} \\
\end{pmatrix}
\label{massmatrix}
\end{eqnarray}
where
\begin{eqnarray}
m_{hh}^{'} & \ = \ & 2v^{2}\lambda_{H},\nonumber\\
m_{hS_{1}}^{'} & \ = \ & vu_{1}\lambda_{H\phi_{1}}=m_{S_{1}h}^{'} , \nonumber\\
m_{hS_{2}}^{'} & \ = \ & vu_{2} \lambda_{H\Phi_{2}}=m_{S_{2}h}^{'} , \nonumber\\
m_{hS_{3}}^{'} & \ = \ & vu_{3} \lambda_{H\Phi_{3}}=m_{S_{3}h}^{'} , \nonumber\\
m_{S_{1}S_{1}}^{'} & \ = \ & \dfrac{4u_{1}^{3}\lambda_{\phi_{1}}-\sqrt{2}u_{2}u_{3}\mu_{\phi}}{2u_{1}} , \nonumber\\
m_{S_{1}S_{2}}^{'} & \ = \ & u_{1}u_{2}\lambda_{\phi_{1}\phi_{2}}+\dfrac{u_{3}\mu_{\phi}}{\sqrt{2}}  =m_{S_{2}S_{1}}^{'}, \nonumber\\
m_{S_{1}S_{3}}^{'} & \ = \ & u_{1}u_{3}\lambda_{\phi_{1}\phi_{3}}+\dfrac{u_{2}\mu_{\phi}}{\sqrt{2}}  =m_{S_{3}S_{1}}^{'}, \nonumber\\
m_{S_{2}S_{2}}^{'} & \ = \ &  \dfrac{4u_{2}^{3}\lambda_{\phi_{2}}-\sqrt{2}u_{1}u_{3}\mu_{\phi}}{2u_{2}}, \nonumber\\
m_{S_{2}S_{3}}^{'} & \ = \ & u_{2}u_{3}\lambda_{\phi_{2}\phi_{3}}+\dfrac{u_{1}\mu_{\phi}}{\sqrt{2}}=m_{S_{3}S_{2}}^{'}, \nonumber\\
m_{S_{3}S_{3}}^{'} & \ = \ & \dfrac{4u_{3}^{3}\lambda_{\phi_{3}}-\sqrt{2}u_{1}u_{2}\mu_{\phi}}{2u_{3}}, \nonumber \\
m_{\eta_{R}\eta_{R}}^{'} & \ = \ & \dfrac{1}{2} \left( v^{2}(\lambda_{H\eta}+\lambda_{H\eta}^{'})+ u_{1}^{2}\lambda_{\eta \phi_{1}}+u_{2}^{2}\lambda_{\eta \phi_{2}}+u_{3}^{2}\lambda_{\eta \phi_{3}}+2\mu_{\eta}^{2} \right) \, .
\label{mass_relation}
\end{eqnarray}
Diagonalising this real symmetric matrix by the orthogonal matrix $\mathcal{O}_{s}$ the physical states can be identified as
\begin{eqnarray}
\begin{pmatrix}
h' \\\eta_{R}' \\S_{1}' \\S_{2}'\\S_{3}'\\ 
\end{pmatrix}
= \mathcal{O}_{s} 
\begin{pmatrix}
h \\ \eta_{R} \\ S_{1} \\S_{2}\\S_{3}\\
\end{pmatrix}.
\end{eqnarray}
where
\begin{eqnarray}
\mathcal{O}_{s}=\mathcal{O}_{12}\mathcal{O}_{13}\mathcal{O}_{14}\mathcal{O}_{15}
\end{eqnarray}
Similarly for the pseudo scalar the $4\times 4$ mass matrix is found to be
\begin{eqnarray}
\begin{pmatrix}
m_{\eta_{I}\eta_{I}}^{'} & 0 & 0 & 0  \\
0 & m_{A_{1}A_{1}}^{'} & m_{A_{1}A_{2}}^{'} & m_{A_{1}A_{3}}^{'}  \\
0  & m'_{A_{2}A_{1}} & m'_{A_{2}A_{2}} & m'_{A_{2}A_{3}} \\
0  & m'_{A_{3}A_{1}} & m'_{A_{3}A_{2}} & m'_{A_{3}A_{3}} \\
\end{pmatrix}
\label{massmatrix}
\end{eqnarray}
where
\begin{eqnarray}
m_{\eta_{I}\eta_{I}}^{'} & \ = \ & \dfrac{1}{2} \left( v^{2}(\lambda_{H\eta}+\lambda_{H\eta}^{'})+u_{1}^{2}\lambda_{\eta \phi_{1}}+u_{2}^{2}\lambda_{\eta \phi_{2}}+u_{3}^{2}\lambda_{\eta \phi_{3}}+\mu_{\eta}^{2} \right),\nonumber\\
m_{A_{1}A_{1}}^{'} & \ = \ &  -\dfrac{u_{2}u_{3}\mu_{\phi}}{\sqrt{2}u_{1}}, \nonumber\\
m_{A_{1}A_{2}}^{'} & \ = \ & \dfrac{u_{3}\mu_{\phi}}{\sqrt{2}}=m_{A_{2}A_{1}}^{'} , \nonumber\\
m_{A_{1}A_{3}}^{'} & \ = \ & -\dfrac{u_{2}\mu_{\phi}}{\sqrt{2}}=m_{A_{3}A_{1}}^{'} , \nonumber\\
m_{A_{2}A_{2}}^{'} & \ = \ & -\dfrac{u_{1}u_{3}\mu_{\phi}}{\sqrt{2}u_{2}}, \nonumber\\
m_{A_{2}A_{3}}^{'} & \ = \ & \dfrac{u_{1}\mu_{\phi}}{\sqrt{2}}  =m_{A_{3}A_{2}}^{'} \, .
\label{mass_relation}
\end{eqnarray}

\section{Relevant cross sections and decay widths}
\label{washouts}

The cross section for the process $N_{1}N_{1}\longrightarrow f \bar{f}$ is found out to be

\begin{align}\label{eq:C1}
\sigma_{N_{1}N_{1}\longrightarrow f\bar{f}} & =\dfrac{n^2_1g_{BL}^{4}}{48\pi \{ (s-M_{Z_{BL}}^{2})^{2}+\Gamma_{Z_{BL}}^{2}M_{Z_{BL}}^{2} \}} \sum_{f}n_{f}^{2}N_{f}^{C}   \sqrt{1-\dfrac{4m_{f}^{2}}{s}} \left(   2m_{f}^{2}+s\right)  ,
\end{align}

where $n_{f} (n_1)$ is the charge of the SM fermion f ($N_{1}$) under the $U(1)_{B-L}$ and $N_{f}^{C}$ is the colour multiplicity of the fermions. \\

The total washout term $W^{\rm Total}$ given in equation \eqref{eq:6} contains multiple washout processes and can be identified to be

\begin{align} \label{eq:C2}
W^{\rm Total} & =W_{\rm ID}+W_{lZ_{BL}\longrightarrow \eta N_{1}}+W_{\eta l \longrightarrow N_{1}Z_{BL}}+W_{l\eta \longleftarrow \bar{l}\eta^{*}}+W_{l W^{\pm} (Z)\longrightarrow \eta N_{1}} \nonumber \\ & W_{l N_{1} \longrightarrow
 \bar{l} N_{1}^{*}}.
\end{align}

Here $W_{\rm ID}$ is the inverse decay term and it is given by

\begin{equation} \label{eq:c3}
W_{\rm ID}=\dfrac{1}{4}K_{N_{1}}z^{3}\kappa_{1}(z_{1}),
\end{equation}

where $K_{N_{1}}=\dfrac{\Gamma_{1}}{H(z=1)}$ is the decay parameter and the corresponding decay width of $N_{1}$ is given by 

\begin{equation} \label{eq:C4}
\Gamma_{1}=\dfrac{M_{1}}{8\pi}\left( Y_{\eta}^{\dagger} Y_{\eta} \right)_{11}\left(  1-\dfrac{m_{\eta}^{2}}{M_{1}^{2}} \right).
\end{equation}

The other relevant washout terms are defined by 

\begin{eqnarray} \label{eq:washouts}
W_{l Z_{BL}\longrightarrow \eta N_{1}} & \ = & \ \dfrac{z s}{H(z=1)} n_{Z_{BL}}^{eq}\langle  \sigma v \rangle_{lZ_{BL}\longleftarrow \eta N_{1}}, \\
W_{l \eta \longleftrightarrow N_{1}Z_{BL}} & \ = & \ \dfrac{z s}{H(z=1)} n_{\eta}^{eq}\langle \sigma v  \rangle_{l \eta \longrightarrow N_{1}Z_{BL}}, \\
W_{l\eta \longrightarrow \bar{l} \eta^{*}} & \ = & \ \dfrac{z s}{H(z=1)}n_{\eta}^{eq} \langle \sigma v \rangle_{l \eta \longrightarrow \bar{l}\eta^{*}}, \\
W_{l W^{\pm}(Z)\longleftarrow \eta N_{1}} & \ = & \ \dfrac{z s}{H(z=1)}n_{W(Z)}^{eq} \langle \sigma v \rangle_{lW^{\pm}(Z)\longrightarrow \eta N_{1}}, \\
W_{l N_{1} \longleftarrow \bar{l} N_{1}^{*}} & \ = & \ \dfrac{z s}{H(z=1)} n_{N_{1}}^{eq} \langle \sigma v \rangle_{l N_{1} \longleftarrow \bar{l} N_{1}^{*}}.
\end{eqnarray}

Where, $\langle \sigma v \rangle_{ij\longrightarrow k,l}$ is the thermal averaged cross section for the process $i,j \longrightarrow k,l$ and is given by 
 
\begin{align}
\langle \sigma v \rangle_{i,j\longrightarrow k,l} & =\dfrac{z}{8m_{i}^{2}m_{j}^{2}K_{2}\left(\dfrac{m_{i}}{M_{1}}z\right)K_{2}\left(\dfrac{m_{j}}{M_{1}}z\right)} \int_{(m_{i}+m_{j})^{2}}^{\infty} \sigma_{ij\longleftarrow kl}(s-(m_{i}+m_{j})^{2}) \nonumber \\ & \sqrt{s}K_{1}(\sqrt{s}z/M_{1}).
\end{align}

The relevant cross sections (assuming $n_1=-3/2$ as in the UV complete model) for the washouts are given below.

\begin{align}
\sigma_{lZ_{BL}\longleftarrow \eta N_{1}} & = \dfrac{Y_{\eta}^{2}g_{BL}^{2}}{8\pi s}  \left[ \dfrac{(s-(m_{\eta}+M_{1})^{2})(s-(m_{\eta}-M_{1})^{2})}{(s-(m_{l}+M_{Z_{BL}})^{2})(s-(m_{l}-M_{Z_{BL}})^{2})}  \right]^{1/2}  \nonumber \\ & \  \dfrac{(m_{l}^{2}+2M_{Z_{BL}}^{2})}{s^{2}} \left(\dfrac{s-M_{Z_{BL}}^{2}}{2\sqrt{s}} \left( \dfrac{s-m_{\eta}^{2}+M_{1}^{2}}{2\sqrt{s}}+\sqrt{\left( \dfrac{s-m_{\eta}^{2}+M_{1}^{2}}{2\sqrt{s}} \right)^{2}-M_{1}^{2}} \right)  \right)
\end{align} 

\begin{align}
\sigma_{l \eta \longrightarrow N_{1} Z_{BL}} & =\dfrac{9Y_{\eta}^{2}g_{BL}^{2}}{32\pi s}  \left[ \dfrac{(s-(M_{1}+M_{Z_{BL}})^{2})(s-(M_{1}-M_{Z_{BL}})^{2})}{(s-(m_{l}+m_{\eta})^{2})(s-(m_{l}-m_{\eta})^{2})}  \right]^{1/2}  \nonumber \\ & \  \dfrac{1}{(s-M_{1}^{2})^{2}}  \left( \dfrac{s+M_{1}^{2}-M_{Z_{BL}}^{2}}{2\sqrt{s}}.\dfrac{s-M_{1}^{2}+M_{Z_{BL}}^{2}}{2\sqrt{s}}+\sqrt{\left( \dfrac{s-m_{\eta}^{2}}{2\sqrt{s}}\right)}\sqrt{\left(\dfrac{s-m_{\eta}^{2}}{2\sqrt{s}} \right)^{2}-M_{1}^{2}} \right) \nonumber \\ & \left(   \dfrac{s-m_{\eta}^{2}}{2\sqrt{s}}\dfrac{s-M_{1}^{2}+M_{Z_{BL}}^{2}}{2\sqrt{s}}+\dfrac{s-m_{\eta}^{2}}{2\sqrt{s}} \sqrt{\left( \dfrac{s-m_{\eta}^{2}}{2\sqrt{s}}\right)^{2}-M_{1}^{2}} \right)
\end{align}

For the processes $\sigma_{l Z_{BL} \longrightarrow N_{1} Z_{BL}}$ and $\sigma_{l\eta \longrightarrow N_{1}Z_{BL}}$, we have written the cross sections coming from the s-channel diagram only as they are the dominant ones in this case. However, in our numerical analysis we have taken the contribution from both s and t channel diagrams and their interferences.

\begin{align}
\sigma_{lW^{\pm}(Z)\longleftarrow \eta N_{1}} & = \dfrac{Y_{\eta}^{2}g^{2}}{64\pi s}  \left[ \dfrac{(s-(m_{\eta}+M_{1})^{2})(s-(m_{\eta}-M_{1})^{2})}{(s-(m_{l}+M_{W})^{2})(s-(m_{l}-M_{W})^{2})}  \right]^{1/2}  \nonumber \\ & \  \dfrac{(m_{l}^{2}+2M_{W}^{2})}{s^{2}} \left(\dfrac{s-M_{W}^{2}}{2\sqrt{s}} \left( \dfrac{s-m_{\eta}^{2}+M_{1}^{2}}{2\sqrt{s}}+\sqrt{\left( \dfrac{s-m_{\eta}^{2}+M_{1}^{2}}{2\sqrt{s}} \right)^{2}-M_{1}^{2}} \right)  \right)
\end{align} 

\begin{align}
\sigma_{l\eta \longrightarrow \bar{l}\eta^{*}} & =\dfrac{Y_{\eta}^{4}}{4\pi s}\dfrac{M_{1}^{2}}{(s-M_{1}^{2})^{2}}\left[ \dfrac{s-m_{\eta}^{2}+m_{l}^{2}}{2\sqrt{s}}\dfrac{s-m_{l}^{2}+m_{\eta}^{2}}{2\sqrt{s}} \right]
\end{align}

\begin{align}
\sigma_{lN_{1}\longrightarrow \bar{l}N_{1}} & = \dfrac{Y_{\eta}^{4}}{4\pi s}\dfrac{m_{\eta}^{2}}{(s-m_{\eta}^{2})^{2}}\left[ \dfrac{s-M_{1}^{2}}{2\sqrt{s}}\dfrac{s-m_{l}^{2}+M_{1}^{2}}{2\sqrt{s}} + \dfrac{s-m_{l}^{2}+M_{1}^{2}}{2\sqrt{s}}\sqrt{\left( \dfrac{s-m_{l}^{2}+M_{1}^{2}}{2\sqrt{s}} \right)^{2}-M_{1}^{2}} \right]
\end{align}

\begin{figure}
\begin{center}
\includegraphics[scale=.5]{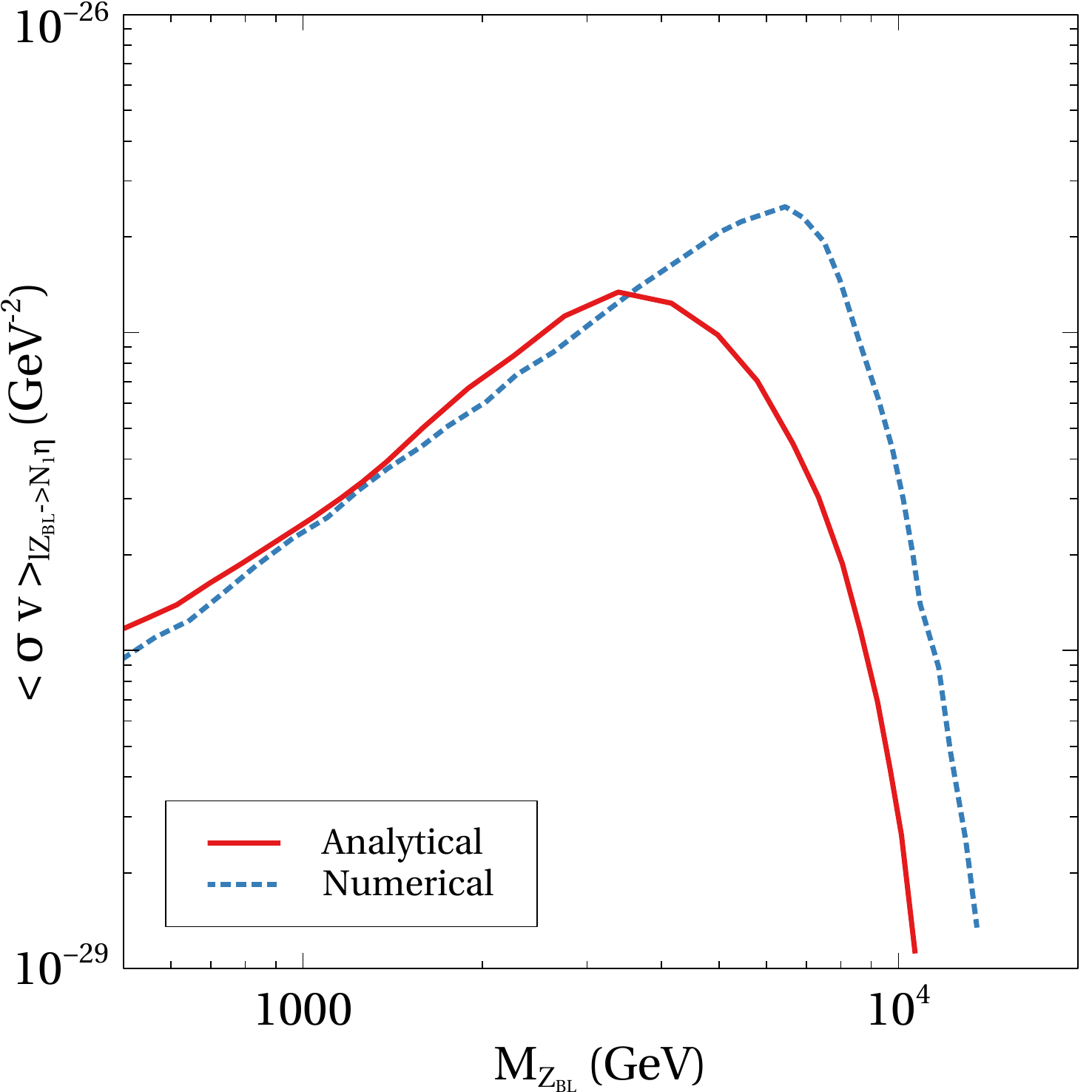}
\caption{Comparison of analytical and numerical results for the thermal averaged cross section of the key washout process $l Z_{BL} \longrightarrow N_{1} \eta$ at temperature $T=100 M_{1}$. The relevant benchmark parameters are fixed at the following values $m_{\eta}=500$ GeV, $M_{1}=10$ TeV, $g_{BL}=0.1$, $(Y_{\eta})_{\alpha 1}\simeq 10^{-4}$. }
\label{appenfig}
\end{center}
\end{figure}

In figure \ref{appenfig}, we show the behaviour of thermal averaged cross section for $l Z_{BL} \longrightarrow N_{1} \eta$, a key washout process involving $B-L$ gauge boson. The analytical estimate matches with the numerical estimates extracted from \texttt{micrOMEGAs} quite well. Clearly, with rise in $Z_{BL}$ mass, the cross section increases enhancing the washout effects, as pointed out in the discussions above. Also, for very heavy $Z_{BL}$ corresponding to a decoupled scenario, there is no impact of $Z_{BL}$ mass on leptogenesis due to tiny washout effects at the scale of leptogenesis.

\begin{figure}
\begin{center}
\includegraphics[scale=.5]{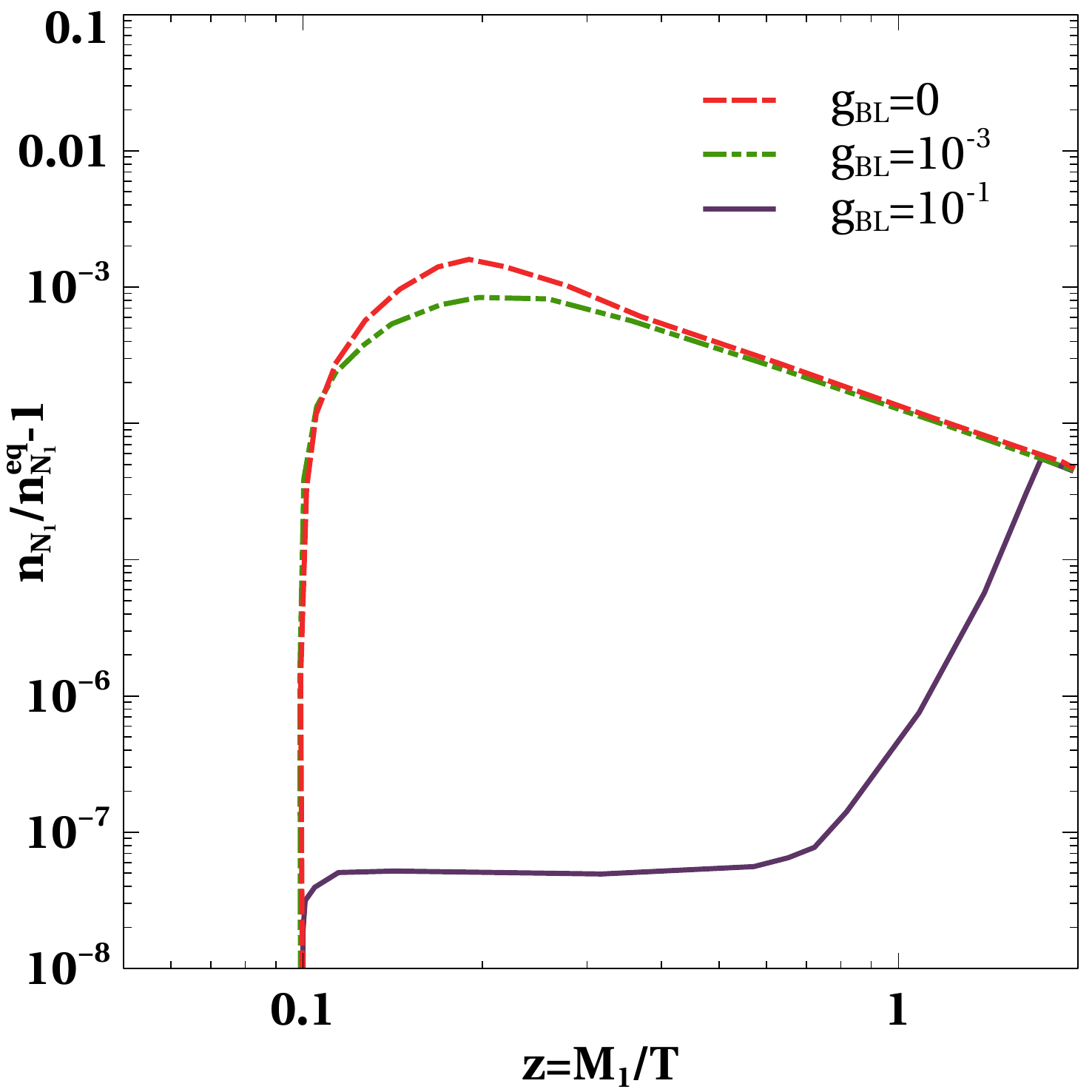}
\includegraphics[scale=.5]{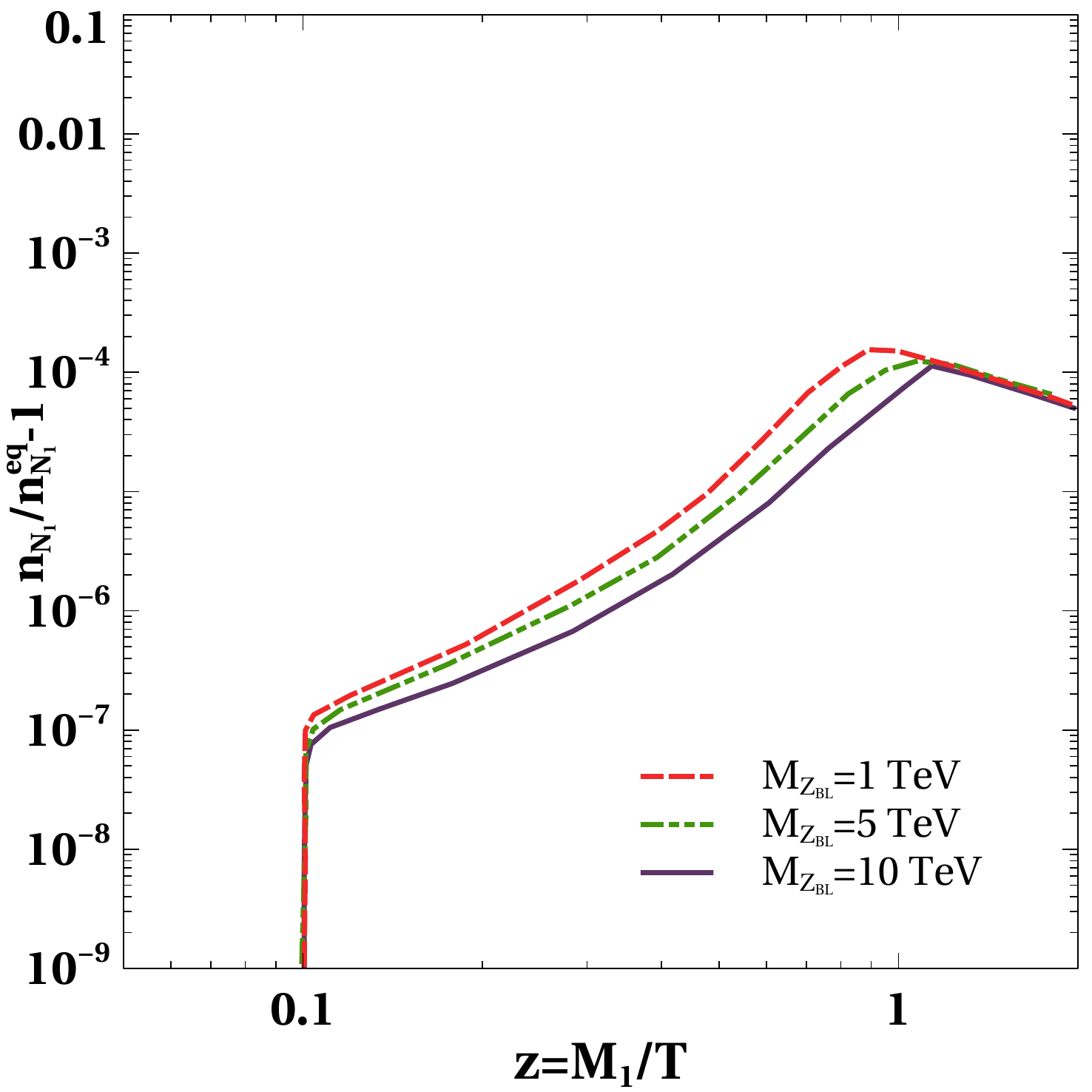}
\caption{The deviation of the comoving number density of $N_1$ (right panel) with $z=\dfrac{M_{1}}{T}$ from its equilibrium density for different $g_{BL}$ (left panel) and for different $M_{Z_{BL}}$ (right panel). The Yukawa couplings relevant for leptogenesis are taken to be $(Y_{\eta})_{\alpha 1}=10^{-5}(1+i)$ and $(Y_{\eta})_{\alpha 2}=10^{-1}(1.1-i)$. The other important parameters used are $M_{1}=15$ TeV, $M_{2}=150$ TeV, $m_{\eta}=5$ TeV. For left (right) panel plot we fix $M_{Z_{BL}}=5$ TeV ($g_{BL}=10^{-3}$). }
\label{equilibrium}
\end{center}
\end{figure}

In figure \ref{equilibrium}, we show the effects of $B-L$ gauge sector parameters namely, gauge coupling $g_{BL}$ and gauge boson mass $M_{Z_{BL}}$ on departure in evolution of comoving number density of $N_1$ from its equilibrium number density. We can clearly notice significant effects due to these parameters as additional gauge interactions lead to $N_{1}$ annihilations keeping it in equilibrium for longer duration. With the increase in $g_{BL}$ the rate of $N_{1}$ annihilations increases which brings $N_{1}$ no density very close to its equilibrium number density, as seen from left panel plot of figure \ref{equilibrium}. Also, similar behaviour can be seen with the increase in $M_{Z_{BL}}$. Since we are working in the parameter space where $M_{Z_{BL}} \le 2M_{1}$, with the increase in $M_{Z_{BL}}$ the cross section for $N_{1}$ annihilation increases which again bring the $N_{1}$ no density closer to its equilibrium number density.

Finally, we write the expression for $\nu_{R}\nu_{R}\longrightarrow f \bar{f}$ cross section mediated by $Z_{BL}$ used in the estimate for $\Delta N_{eff}$. This is given by

\begin{align}
\sigma_{\nu_{R}\nu_{R}\longrightarrow f \bar{f}}=\dfrac{g_{BL}^{4}}{48\pi}\dfrac{1}{(s-M_{Z_{BL}}^{2})+\Gamma_{Z_{BL}}^{2}M_{Z_{BL}}^{2}}\sum_{f}n_{f}^{2}N_{f}^{C}\sqrt{1-\frac{4m_{f}^{2}}{s}}\left( s+2m_{f}^{2}\right).
\end{align}

\section{CP asymmetry}
\label{asymmetry}
The decay width for the decay $N_{1} \longrightarrow \eta l$ is given by 
\begin{equation}
  \Gamma_{N_{1}\longrightarrow \eta l} =\dfrac{M_{1}}{8 \pi}(Y^{\dagger}_{\eta} Y_{\eta})_{11}\left(1-\dfrac{m_{\eta}^2}{M_{1}^2}\right)^2
\end{equation}
The CP asymmetry parameter for $N_1 \rightarrow l_{\alpha} \eta, \bar{l_{\alpha}}\bar{\eta}$ is given by 
\begin{align}
\epsilon_{1 \alpha} & = \frac{1}{8 \pi (Y^{\dagger}_{\eta} Y_{\eta})_{11}} \bigg [ f \left( \frac{M^2_2}{M^2_1}, \frac{m^2_{\eta}}{M^2_1} \right) {\rm Im} [ (Y_{\eta})^*_{\alpha 1} (Y_{\eta})_{\alpha 2} (Y^{\dagger}_{\eta} Y_{\eta})_{12}] \nonumber \\
& -\frac{M^2_1}{M^2_2-M^2_1} \left( 1-\frac{m^2_{\eta}}{M^2_1} \right)^2 {\rm Im}[(Y_{\eta})^*_{\alpha 1} (Y_{\eta})_{\alpha 2} H_{12}] \bigg ]
\label{epsilonflav}
\end{align}
where, the function $f(r_{ji},\eta_{i})$ is coming from the interference of the tree-level and one loop diagrams and has the form
\begin{equation}
f(r_{ji},\eta_{i})=\sqrt{r_{ji}}\left[1+\frac{(1-2\eta_{i}+r_{ji})}{(1-\eta_{i}^{2})^{2}}{\rm ln}(\frac{r_{ji}-\eta_{i}^{2}}{1-2\eta_{i}+r_{ji}})\right]
\end{equation}
with $r_{ji}=M_{j}^{2}/M_{i}^{2}$ and $\eta_{i}=m_{\eta}^{2}/M_{i}^{2}$. The self energy contribution $H_{ij}$ is given by 
\begin{equation}
H_{ij} = (Y^{\dagger}_{\eta} Y_{\eta})_{ij} \frac{M_j}{M_i} + (Y^{\dagger}_{\eta} Y_{\eta})^*_{ij}
\end{equation}
Now, the CP asymmetry parameter, neglecting the flavour effects (summing over final state flavours $\alpha$) is
\begin{equation}
\epsilon_{1}=\frac{1}{8\pi(Y^{\dagger}_{\eta} Y_{\eta})_{11}} {\rm Im}[((Y^{\dagger}_{\eta} Y_{\eta})_{12})^{2}]\frac{1}{\sqrt{r_{21}}}F(r_{21},\eta_{1})
 \label{eq:14}
\end{equation}
\\
where the function $F(r_{ji},\eta)$ is defined as 
\begin{equation}
F(r_{ji},\eta_{i})=\sqrt{r_{ji}}\left[ f(r_{ji},\eta_{i})-\frac{\sqrt{r_{ji}}}{r_{ji}-1}(1-\eta_{i})^{2} \right].
\end{equation}
Let us denote the Yukawa couplings as
\begin{eqnarray}
 (Y_{\eta})_{\alpha 1} & \ = & \ a+i a, \\
 (Y_{\eta})_{\alpha 2} & \ = & \ b-i c,
\end{eqnarray}
where $a,b,c$ are three real numbers. For this choice of Yukawa, The CP asymmetry parameter for $N_{1}$ decay is 
\begin{eqnarray}
 \epsilon_1  & \ \propto & \ \dfrac{{\rm Im}[(Y_{\eta}^{\dagger}Y_{\eta})_{12}^{2}]}{(Y_{\eta}^{\dagger}Y_{\eta})_{11}} \nonumber \\
           & = & \dfrac{-18a^{2}({\rm Re}[(b-ic)^{2}])}{6a^{2}} \nonumber \\
           & = & -3({\rm Re}[(b-ic)^{2}]) = -3 (b^2-c^2)
\end{eqnarray}
Here one can also notice that to have a non-vanishing CP asymmetry we must have $b \ne c$.

\providecommand{\href}[2]{#2}\begingroup\raggedright\endgroup

\end{document}